\documentclass{article}
\usepackage{newtxtext}

\usepackage{arxiv}

\usepackage[utf8]{inputenc} 
\usepackage[T1]{fontenc}    
\usepackage{hyperref}       
\usepackage{url}            
\usepackage{booktabs}       
\usepackage{amsfonts}       
\usepackage{nicefrac}       
\usepackage{microtype}      

\usepackage{subfig}         
\usepackage{amsmath}        
\usepackage{listings}       
\usepackage{multirow}       
\usepackage{lipsum}		
\usepackage{graphicx}
\usepackage{natbib}
\usepackage{doi}

\title{Beyond Accuracy: LLM Variability in Evidence Screening for Software Engineering SLRs}

\author{ \href{https://orcid.org/0009-0003-2310-5323}{\includegraphics[scale=0.06]{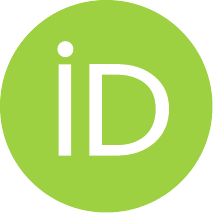}\hspace{1mm}Gilberto Sussumu Hida}\thanks{Email: sussumu.educacional@gmail.com, } \\
	AIBL\\
	CESAR School\\
	Recife, Pernambuco, Brazil \\
	\texttt{gsh@cesar.school} \\
	\And
	\href{https://orcid.org/0009-0003-7032-8167}{\includegraphics[scale=0.06]{orcid.pdf}\hspace{1mm}Danilo Monteiro Ribeiro} \\
	AIBL\\
	CESAR School\\
	Recife, Pernambuco, Brazil \\
	\texttt{dmr@cesar.school} \\
    \and
    \href{https://orcid.org/ 0009-0007-8531-8446}{\includegraphics[scale=0.06]{orcid.pdf}\hspace{1mm}Erika Yahata} \\
	CMCC\\
	UFABC\\
	Santo André, São Paulo, Brazil \\
	\texttt{e.yahata@ufabc.edu.br} \\
}

\renewcommand{\shorttitle}{Beyond Accuracy: LLM Variability in SLR Screening}

\hypersetup{
pdftitle={Beyond Accuracy: LLM Variability in Evidence Screening for Software Engineering SLRs},
pdfsubject={cs.SE, cs.AI, cs.LG},
pdfauthor={Hida. Gilberto Sussumu; Ribeiro.~Danilo Monteiro; Yahata.~Erika},
pdfkeywords={Systematic Literature Review, Large Language Models, LLMs, Automated Screening, Machine Learning, Feature Engineering, Software Engineering},
}

\begin{document}
\maketitle

\begin{abstract}
\textbf{Context:} Study screening in systematic literature reviews is costly, inconsistency-prone, and risk-asymmetric, since false negatives can compromise validity. Despite rapid uptake of Large Language Models (LLMs), there is limited evidence on how such models behave during the study screening phase, particularly regarding the choice of specific LLMs and their comparison with classical models. \textbf{Objective:} To assess LLM performance and variability in screening, quantify the impact of input metadata (abstract, title, keywords), and compare LLMs with classical classifiers under a shared protocol.
\textbf{Methods:} We analyzed 12 LLMs from 4 providers (OpenAI, Google Gemini, Anthropic, Llama) and 4 classical models (Logistic Regression, Support Vector Classification, Random Forest, and Naive Bayes) on 2 real Systematic Literature Reviews (SLRs), totaling 518 papers. The experimental design investigated 3 critical dimensions: (i) LLMs performance variability, (ii) the impact of input feature composition (abstract, title, and keywords) on LLM performance, and (iii) the real gain of using LLMs instead of more traditional classification models. \textbf{Results:} LLMs exhibited substantial heterogeneity and residual non-determinism even at temperature zero. Abstract availability was decisive: removing it consistently degraded performance, while adding title and/or keywords to the abstract yielded no robust gains. Compared to classical models, performance differences were not consistent enough to support generalizable LLM superiority. \textbf{Discussion:} LLM adoption should be justified by operational and governance constraints (reproducibility, cost, metadata availability), supported by pilot validation and explicit reporting of variability and input configuration.
\end{abstract}


\begin{quote}
\small
\textbf{Preprint note.} This manuscript is an earlier, shorter, conference-style version of a more comprehensive journal manuscript currently under review. The journal submission is substantially expanded and incorporates revisions motivated by prior peer review, as well as additional methodological details, limitations, and practical guidance.
\end{quote}

\maketitle

\section{Introduction}
Systematic Literature Reviews (SLRs) constitute an essential approach for consolidating knowledge in Software Engineering (SE)~\cite{costalongaCanMachineLearning2025a, felizardoChatGPTApplicationSystematic2024a, syrianiAssessingAbilityChatGPTb}, enabling the rigorous synthesis of evidence across diverse topics in the field. However, the process can be time-consuming, costly, and susceptible to bias~\cite{vandinterAutomationSystematicLiterature2021}, particularly during the publication screening phase, which involves analyzing titles, abstracts, and keywords. This step is vulnerable to fatigue, confirmation bias, and inconsistencies across reviewers, which limits scalability when the corpus reaches thousands of records. Established guidelines, such as involving independent pairs of reviewers followed by cross-validation, help mitigate these risks but do not eliminate the high operational cost of the process~\cite{costalongaCanMachineLearning2025a, felizardoChatGPTApplicationSystematic2024a, thodeExploringUseLLMs2025a, wangErrorRatesHuman2020a}.

To reduce this cost, the scientific community has explored alternatives to support study selection~\cite{espositoGenerativeAIEvidenceBased2024, felizardoChatGPTApplicationSystematic2024a, luo2024potential}. Supervised machine learning approaches have demonstrated the potential to reduce human effort~\cite{minettonapoleaoEmergingResultsAutomated2024}, and among these approaches, LLMs have emerged as promising candidates due to their capacity for deep contextual understanding~\cite{brownLanguageModelsAre2020}. 

Because they are emerging technologies, LLMs present operational and behavioral challenges that are not yet fully understood. Recent studies have reported issues related to stability, cost and reproducibility~\cite{thodeExploringUseLLMs2025a, felizardoChatGPTApplicationSystematic2024a}, especially when models are accessed through commercial APIs. Research has also identified the phenomenon of hallucination, which, in the present context, corresponds to the generation of non-existent references or fabricated information~\cite{huangSurveyHallucinationLarge2025, xuHallucinationInevitableInnate2025, rawteSurveyHallucinationLarge2023}. Although these issues are more evident in the synthesis stages, they may also affect reliability in the early screening phases. These characteristics, combined with the black-box nature of the models~\cite{bender2021dangers} and the rapid evolution of versions~\cite{chen2024chatgpt}, indicate that the overall behavior of LLMs in scientific research contexts still requires systematic investigations to establish their scope, limits, and appropriate conditions of use~\cite{angermeir2025reflections}.

In this context, we conducted a controlled empirical evaluation comparing 12 LLMs from 4 providers and 4 classical classifiers in the screening task on 2 real SLRs, previously used in the related literature~\cite{felizardoChatGPTApplicationSystematic2024a}. The study was designed to produce evidence for the following research questions: (RQ1) how performance and variability behave between models and iterations; (RQ2) the impact of metadata compositions (with and without abstracts); and (RQ3) whether there is a consistent predictive gain compared to traditional approaches.

\section{Related Work}

Automation of the screening phase in SLRs has received increasing attention from the community, particularly with the emergence of Large Language Models (LLMs). This section synthesizes studies that investigate the use of LLMs and classical Machine Learning techniques to support study selection in Software Engineering.

\subsection{Classical Machine Learning Approaches}
Classical Machine Learning approaches to screening generally depend on labeled data and a training and validation process, enabling explicit optimization toward objectives such as high recall, which is often desirable in review screening~\cite{cohen2006reducing}. 

~\cite{costalongaCanMachineLearning2025a} investigated the specific context of SLR updates, training Random Forest (RF) and Support Vector Machines (SVM) on data from the original SLR for application to 551 new studies. RF achieved an F-score of 0.33, while the SVM optimized for high recall achieved 100\% recall with 10\% precision, enabling the exclusion of 33.9\% of studies without loss of evidence. A relevant finding was that human--human pairs produced results significantly more aligned with the final curated outcome than human--ML pairs~\cite{costalongaCanMachineLearning2025a}.

~\cite{minettonapoleaoEmergingResultsAutomated2024} proposed an approach that automates both search (via snowballing) and study selection. In the search phase, the system identified 95.3\% of the studies retrieved manually. For selection, among 4 algorithms tested (LSVM, XGBoost, Logistic Regression, Multinomial Naive Bayes), LSVM presented the best performance (recall 74.3\%, precision 15\%, F-measure 0.246), reducing by approximately 2.5 times the number of studies that require manual analysis~\cite{minettonapoleaoEmergingResultsAutomated2024}.

\subsection{LLMs for Study Screening}

Recent studies investigate LLMs for SLR screening using title, abstract, and keywords as input. A critical gap is that most of the studies treat metadata as a single input without systematically isolating the effect of different compositions on performance.

\begin{figure}[!t]
  \centering
  \includegraphics[width=0.45\textwidth]{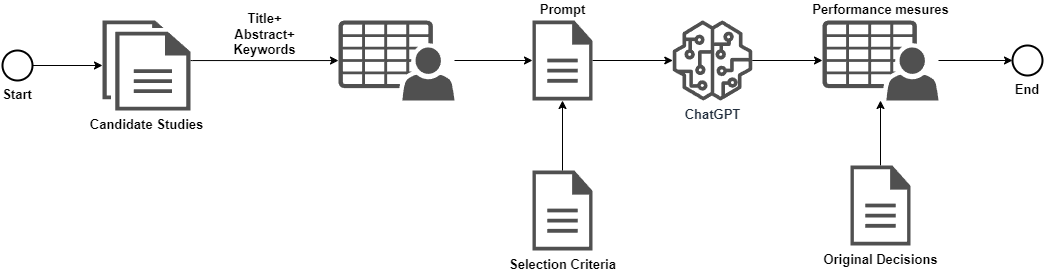}
  \caption{Methodological structure of the study by ~\cite{felizardoChatGPTApplicationSystematic2024a}}
  \label{diagrama_estudo_katia}
\end{figure}

~\cite{felizardoChatGPTApplicationSystematic2024a} conducted a systematic replication using ChatGPT-4.0 on 2 SE SLRs. Figure~\ref{diagrama_estudo_katia} illustrates the methodological structure.

~\cite{thodeExploringUseLLMs2025a} evaluated 5 LLM variants (Llama 2, GPT-3.5/4-turbo, Mixtral) on 2 SE datasets, adopting recall maximization. Combining LLMs via unanimity achieved 99\% recall (27\% precision), suggesting effort reduction without evidence loss.

~\cite{syrianiAssessingAbilityChatGPTb} compared ChatGPT and 4 classical classifiers on 5 ReLiS datasets (5,144 studies). ChatGPT achieved mean recall 0.74 (range: 0.33--0.95), comparable to classical methods (0.67--0.73). While balanced accuracy and MCC were similar, ChatGPT required no prior training. Performance varied by dataset characteristics: high recall in typical SLRs (10\% inclusion), balanced in balanced datasets, and degraded in high-conflict datasets ($>$50\%)~\cite{syrianiAssessingAbilityChatGPTb}.

\subsection{Gaps and Positioning of the Present Study}
 
Although existing studies highlight LLM and classical method potential, they differ substantially in metrics, decision strategies, and experimental configurations, limiting generalization. Critical gaps remain: comparisons under identical protocols are scarce, metadata composition effects are not evaluated systematically, and intra-model variability under deterministic configurations is seldom discussed. This study addresses these gaps through controlled  comparative evaluation of performance, metadata influence, and response  consistency.

\section{Methods}
\subsection{Study Design and Research Questions}

This study extends ~\cite{felizardoChatGPTApplicationSystematic2024a} by evaluating multiple API-accessible LLMs under a controlled protocol, quantifying intra-model variability, testing input metadata compositions, and comparing against classical supervised baselines. We address 3 research questions:

\begin{itemize}
    \item \label{RQ1} \textbf{RQ1 - LLM Performance and Variability in Screening:} What are the performance differences across LLM families (OpenAI, Gemini, Claude, Llama) in screening, and how stable are decisions across repeated runs?
    \item \label{RQ2} \textbf{RQ2 - Impact of Input Features:} Which textual metadata composition (title, abstract, keywords) yields the best screening performance, and does the effect vary across models?
    \item \label{RQ3} \textbf{RQ3 - Comparison: LLMs versus Classical Approaches:} Do LLMs provide advantages over classical Machine Learning methods for SLR screening?
\end{itemize}

\begin{figure*}[!t]
  \centering
  \includegraphics[width=0.9\textwidth]{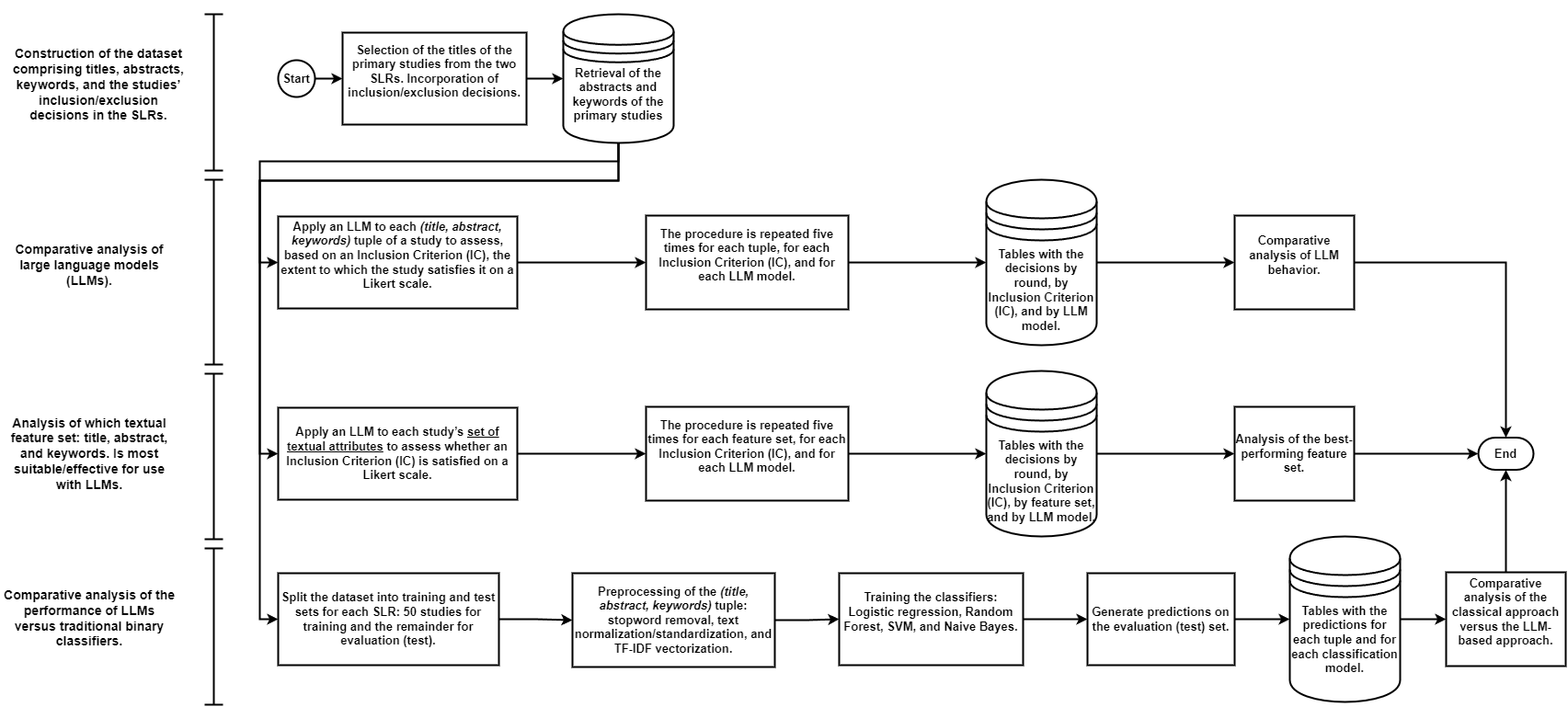} 
  \caption{Experimental pipeline: Data preparation; Comparative analysis of LLM performance; Identification of the most relevant features; Comparison with the traditional classifier-training approach.}
  \label{fig:pipeline_metodologico}
\end{figure*}

We used the same records and reference labels as ~\cite{felizardoChatGPTApplicationSystematic2024a}, restricted to studies with consistently recovered title, abstract, and keywords. Figure~\ref{fig:pipeline_metodologico} summarizes the experimental pipeline across dataset preparation, model execution, and statistical analysis.

The subsequent sections describe each of these phases in detail.

\subsection{Phase 0. Dataset Preparation and Construction}

\textbf{SLR 1:} We reused the SLR1 sample from ~\cite{felizardoChatGPTApplicationSystematic2024a}, which was originally drawn from a tertiary study on HCI--AI convergence~\cite{travassosTertiaryStudyConvergence2017}. From the 134 sampled papers, we extracted the title, abstract, keywords, and the original inclusion/exclusion decisions.

\textbf{SLR 2:} We reused the SLR2 candidate set from a systematic mapping study on user classification strategies in game-based and gamified learning environments~\cite{pessoaJourneyIdentifyUsers2024}, conducted following~\cite{kitchenham2004evidence}. The original search retrieved 579 records; after deduplication, 448 unique candidates remained and served as input to our analyses.

Inclusion criteria (IC) are described in Table~\ref{tab:criterios_ic_rsls}.

\begin{table}[!t]
\centering
\caption{Inclusion criteria for the SLRs}
\label{tab:criterios_ic_rsls}
\begin{tabular}{@{}lp{6cm}@{}}
\toprule
\textbf{SLR} & \textbf{Inclusion Criteria} \\ 
\midrule
\multirow{2}{*}{SLR 1~\cite{travassosTertiaryStudyConvergence2017}} 
& 1. It is a secondary study (Systematic Review, Mapping Study, Rapid Review, or Systematic Mapping) \\
\cmidrule(lr){2-2}
& 2. It presents findings for converging Human-Computer Interaction and Artificial Intelligence \\  \midrule
\addlinespace
\multirow{2}{*}{SLR 2~\cite{pessoaJourneyIdentifyUsers2024}} 
& 1. The publication should discuss the relationship between game elements and user types. \\
\cmidrule(lr){2-2}
& 2. The publication should discuss how to use the relationship between game elements and user types to analyze user engagement. \\
\bottomrule
\end{tabular}
\end{table}

\subsection{Phase 1: Comparative Analysis across LLMs}

This phase extended the investigation by ~\cite{felizardoChatGPTApplicationSystematic2024a}, replicating the study screening task but with multiple LLMs from different providers. Although the original study evaluated a single model (ChatGPT-4.0), in this phase, we evaluated 12 LLMs organized into 4 variants per provider:
\begin{itemize}
    \item \textbf{~\cite{openaiGPTModels2025}}: gpt-3.5-turbo, gpt-4.1, and gpt-4o.
    \item \textbf{Google~\cite{googleGeminiModels2025}}: gemini-1.5-flash, gemini-2.0-flash, and gemini-2.5-flash.
    \item \textbf{~\cite{anthropicClaudeModels2025}}: claude-3-haiku-20240307, claude-3-5-haiku-20241022, and claude-3-5-sonnet-20241022.
    \item \textbf{Llama~\cite{togetherai2025}}: Llama-3.2-3B-Instruct-Turbo, Llama-3.3-70B-Instruct-Turbo, and Llama-4-Scout-17B-16E-Instruct.
\end{itemize}

\begin{figure*}[!t]
  \centering
  \includegraphics[width=0.9\textwidth]{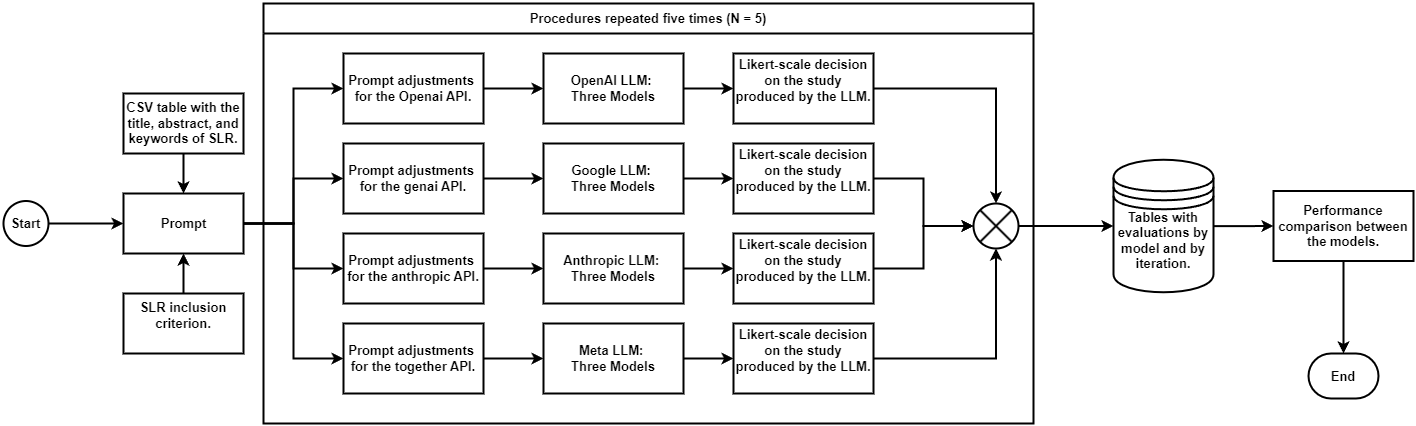} 
  \caption{Phase 1 pipeline for model comparison.}
  \label{fig:pipeline_fase2}
\end{figure*}

The experimental procedure of this phase was structured into 3 main steps (Figure~\ref{fig:pipeline_fase2}): (i) prompt construction, (ii) experimental execution with multiple rounds of LLM inference, and (iii) post-processing and performance comparison.

In the prompt construction step, we developed templates instructing each LLM to classify studies using a 7-point Likert scale (1 = Strongly Disagree, 7 = Strongly Agree), following the approach of ~\cite{felizardoChatGPTApplicationSystematic2024a}. For each study (represented by the metadata tuple: title, abstract, and keywords), the prompt asks the model to rate its agreement with one of the 2 inclusion criteria specific to each SLR, as presented in the previous sections (the process is repeated for each inclusion criterion). The prompts were kept constant across all models to ensure comparability, with the temperature set to zero. Figure~\ref{cod:prompt} presents the template used.
 
\begin{figure}[!t]
    \centering
    \begin{lstlisting}
prompt = f"""
Assume you are a software engineering researcher 
conducting a systematic literature review (SLR).

Considering the title, abstract, and keywords of a 
primary study:

Using a Likert scale from 1 to 7 (1 - Strongly Disagree, 
2 - Disagree, 3 - Partially Disagree, 4 - Neutral, 
5 - Partially Agree, 6 - Agree, and 7 - Strongly Agree), 
evaluate your agreement with the following question:

"{inclusion_criteria_question}"

About the paper:
**Title:** {title}
**Abstract:** {abstract}
**Keywords:** {keywords}

Return only a number from 1 to 7, with no additional 
explanation.
"""
    \end{lstlisting}
    \caption{Prompt Template for LLM-based Classification.}
    \label{cod:prompt}
\end{figure}

Each study was independently assessed by each LLM for each review inclusion criterion, producing 2 Likert scores on the 7-point scale, one for each criterion. We then applied the same conversion rule adopted by ~\cite{felizardoChatGPTApplicationSystematic2024a}: a study was classified as included in the selection only when the LLM assigned score was greater than or equal to 5 for both inclusion criteria; otherwise, the study was classified as excluded. Thus, the final decision for each study was obtained by combining the 2 independent assessments.

The experimental execution step consisted of 5 successive iterations of each model in each study in both datasets (SLR1 and SLR2); that is, each task was repeated 5 times. Multiple rounds were necessary to capture the variability inherent in language models. Although the temperature was set to zero with the aim of inducing more deterministic behavior, the experimental design included repeated executions based on recent investigations that challenge the guarantee of full determinism in LLMs~\cite{atilNonDeterminismDeterministicLLM2025, renzeEffectSamplingTemperature2024}. These studies show that determinism in LLMs is not guaranteed by zero temperature due to factors such as race conditions in parallel processing architectures and floating-point rounding in softmax computations. Accordingly, the multiple-iteration design enabled an empirical investigation of the degree of variability in each model's responses, an issue that remains underexplored in the literature on automated SLR screening.
The performance of each model was evaluated using accuracy as the primary metric; we retained this metric because the class imbalance observed in both datasets was low. If the imbalance were high, we would have opted for a different metric~\cite{hidaOverviewMachineLearning2026}. Finally, we assessed agreement across the multiple iterations of each model using Gwet's AC2 coefficient~\cite{gwet2014handbook} with quadratic weighting (for ordinal data on the 1--7 Likert scale). AC2 was chosen over Cohen's Kappa due to its robustness in class-imbalanced settings (typical in SLR screening where inclusion rates are low), avoiding the prevalence paradox that affects the Kappa coefficient.

\subsection{Phase 2: Impact Analysis of Input Features on LLMs}
\label{subsec:metodosfase2}
While Phase 1 evaluated the performance of multiple LLMs using the full metadata composition (title, abstract, and keywords), the present phase systematically investigates how different input feature compositions affect classification quality. To answer this question (RQ2), we selected random samples of 50 studies from each SLR, totaling 100 studies for the feature analysis. These same 100 studies were processed with 5 distinct feature composition variants, in addition to the composition already processed in the previous phase:

\begin{itemize}
    \item \textbf{Variant A}: Abstract only.
    \item \textbf{Variant B}: Abstract + Title.
    \item \textbf{Variant C}: Abstract + Title + Keywords (already processed in the previous phase).
    \item \textbf{Variant D}: Abstract + Keywords.
    \item \textbf{Variant E}: Title + Keywords.
\end{itemize}

The selection of these variants followed a more empirical logic, directly testing the impact of feature sets on performance and enabling the isolation of the incremental contribution of each textual component. \textbf{Variant A} served as the \textbf{reference baseline} against which we directly compared the performance of the simplest composition with the other compositions.

The experimental procedure for this phase followed a structure similar to Phase 1, as shown in Figure~\ref{fig:pipeline_fase2}, with the difference that the processing was replicated for each of the 4 variants of features. The procedure was structured into 2 main steps: the first focused on constructing prompts adapted to each feature variant, and the second on experimental execution with multiple rounds.

In the prompt construction step, we adapted the template used in Phase 1 (shown in Figure~\ref{cod:prompt}) for each of the 4 variants, keeping the remainder of the prompt identical to ensure comparability. Each prompt was dynamically instantiated with the metadata corresponding to the variant under analysis, preserving the 7-point Likert scale and the SLR-specific inclusion criteria.

For statistical analysis, we constructed 95\% confidence intervals by bootstrapping~\cite{tibshirani1993introduction} over the 5 iterations of each model per variant, allowing us to investigate whether performance differences between variants are statistically significant.

To investigate the differential impact of feature compositions, we performed a contrast meta-analysis. The variance in the model was derived from the 95\% confidence intervals obtained by bootstrap stratified with papers~\cite{tibshirani1993introduction}, and the effects were aggregated using the DerSimonian--Laird random-effects model~\cite{dersimonianMetaAnalysisClinicalTrials2015}, providing an estimate of the mean effect, the quantification of heterogeneity (\(I^2\)) and the prediction intervals. Practical relevance was further contextualized via the Smallest Effect Size of Interest (SESOI)~\cite{anvari2021using}, set to \(\pm 2.0\) percentage points of accuracy.

\subsection{Phase 3: Comparison between LLMs and Classical Machine Learning}

This phase complements the previous ones by comparing LLMs against classical supervised classification methods. The objective is to address RQ3, defined in Section ~\ref{RQ3}. To enable this comparison, classical models were trained using subsets of each SLR. We randomly selected 50 studies from each SLR (SLR1 and SLR2) to compose the training set, keeping the remaining studies for testing. The choice of a reduced training set aims to reflect a pragmatic scenario in which only a fraction of the corpus has been initially screened and labeled by humans, allowing us to evaluate the performance of classical classifiers under label constraints and to compare this setting with the zero-shot learning use of LLMs.

We concatenated title, abstract, and keywords and applied TF--IDF vectorization~\cite{saltonTermWeightingAutomatic1988} after standard preprocessing with NLTK~\cite{bird2009natural}. We trained Multinomial Naive Bayes, Logistic Regression, Random Forest, and SVM~\cite{bishop2006pattern, hastie2009elements} using scikit-learn~\cite{pedregosa2011scikit} with standard settings and stratified 4-fold cross-validation. We evaluated accuracy and F1-score~\cite{hidaOverviewMachineLearning2026} and estimated 95\% confidence intervals via bootstrapping.

\section{Results}

This section reports the empirical findings for each experimental phase.

\subsection{Phase 0. Dataset Preparation and Construction}

Starting from the titles provided by Felizardo et al.~\cite{felizardoChatGPTApplicationSystematic2024a}, we retrieved missing abstracts and keywords to enable the analyses in Phases 1--3.

For SLR1, from 134 titles, we retrieved 126 complete records (94\% retrieval rate). Seven papers were excluded due to missing keywords and one paper was not found, resulting in 126 studies. 

For SLR2, from 448 titles, we consolidated 392 complete records (87.5\% retrieval rate). Fifty-six papers were excluded due to duplication, missing abstracts, or inconsistent metadata.

\subsection{Phase 1: LLM Performance and Variability}
\label{subsec:resultados_fase1}

\textbf{Aggregated Accuracy and Performance Variation:} Table~\ref{tab:resultados_acuracia_fase1} reports the aggregated accuracy (mean across the 5 rounds) for each LLM, stratified by dataset (SLR1 and SLR2).
The results evidenced relevant inter-model heterogeneity: in SLR1, gpt-4o achieved the highest mean accuracy $(0.83 \pm 0.03)$, while claude-3-haiku-20240307 and Llama-3.2-3B-Instruct-Turbo presented the lowest values (0.61); in SLR2, gpt-4.1 achieved the highest accuracy $(0.84\pm 0.05)$, while gpt-3.5-turbo presented the lowest performance $(0.37\pm0.03)$. To characterize the dispersion observed across rounds, we also report the standard deviation of the 5 executions and the range corresponding to the mean standard deviation $\pm$, as indicated in Table~\ref{tab:resultados_acuracia_fase1}, avoiding interpreting this range as a confidence interval. Differences of 22 percentage points (p.p.) for SLR1 and 47 percentage points for SLR2 demonstrate this performance heterogeneity between models in the classification task.

\begin{table*}[!t]
\centering
\caption{Phase 1 results. $\text{lim\_{inf}} = mean-std$ and $\text{lim\_{sup}} = mean+std$}
\label{tab:resultados_acuracia_fase1}
\begin{tabular}{@{}clccccc@{}}
\toprule
\textbf{SLR} & \textbf{Model} & \textbf{Accuracy-Mean} & \textbf{Accuracy-Median} & \textbf{Std} & \textbf{lim\_inf} & \textbf{lim\_sup}\\ 
\midrule
1 & gpt-4o & 0.830159 & 0.825397 & 0.026679 & 0.803480 & 0.856838\\
1 & gemini-1.5-flash & 0.815873 & 0.817460 & 0.006640 & 0.809233 & 0.822513\\
1 & Llama-4-Scout-17B-16E-Instruct & 0.803175 & 0.801587 & 0.003549 & 0.799625 & 0.806724\\
1 & gpt-4.1 & 0.779365 & 0.777778 & 0.006640 & 0.772725 & 0.786005 \\
1 & Llama-3.3-70B-Instruct-Turbo-Free & 0.753968 & 0.753968 & 0.000000 & 0.753968 & 0.753968 \\
1 & claude-3-5-haiku-20241022 & 0.752381 & 0.753968 & 0.003549 & 0.748832 & 0.755930 \\
1 & gemini-2.5-flash & 0.746032 & 0.746032 & 0.021735 & 0.724297 & 0.767767 \\
1 & gemini-2.0-flash & 0.730159 & 0.730159 & 0.007937 & 0.722222 & 0.738095 \\
1 & claude-3-5-sonnet-20241022 & 0.698413 & 0.698413 & 0.000000 & 0.698413 & 0.698413 \\
1 & gpt-3.5-turbo & 0.652381 & 0.634921 & 0.035670 & 0.616711 & 0.688051 \\
1 & Llama-3.2-3B-Instruct-Turbo & 0.611111 & 0.611111 & 0.000000 & 0.611111 & 0.611111 \\
1 & claude-3-haiku-20240307 & 0.611111 & 0.611111 & 0.000000 & 0.611111 & 0.611111 \\ 
2 & gpt-4.1 & 0.835204 & 0.834184 & 0.005290 & 0.829914 & 0.840494 \\
2 & claude-3-5-sonnet-20241022 & 0.834694 & 0.834184 & 0.001141 & 0.833553 & 0.835835\\
2 & gpt-4o & 0.830612 & 0.831633 & 0.005290 & 0.825322 & 0.835902\\
2 & claude-3-haiku-20240307 & 0.807143 & 0.806122 & 0.001397 & 0.805746 & 0.808540 \\
2 & claude-3-5-haiku-20241022 & 0.799490 & 0.798469	& 0.001397 & 0.798093 & 0.800887 \\
2 & gemini-2.5-flash & 0.779592 & 0.778061 & 0.011773 & 0.767818 & 0.791365\\
2 & gemini-1.5-flash & 0.767347 & 0.767857 & 0.002795 &	0.764552 & 0.770141\\
2 & gemini-2.0-flash & 0.745408 & 0.742347 & 0.005530 &	0.739878 & 0.750939\\
2 & Llama-4-Scout-17B-16E-Instruct & 0.678061 & 0.678571 & 0.001141 & 0.676920 & 0.679202 \\
2 & Llama-3.2-3B-Instruct-Turbo & 0.676020 & 0.676020 & 0.000000 & 0.676020 & 0.676020\\
2 & Llama-3.3-70B-Instruct-Turbo-Free & 0.668367 & 0.668367 & 0.000000 & 0.668367 & 0.668367  \\
2 & gpt-3.5-turbo & 0.369898 & 0.369898 & 0.003124 & 0.366774 & 0.373022 \\ \bottomrule
\end{tabular}
\end{table*}

\textbf{Performance and Class Imbalance:} Performance on SLR2 was slightly higher than on SLR1 for most models, suggesting that dataset characteristics (inclusion/exclusion proportion, term distribution) may influence the quality of LLM decisions. 
In screening tasks, class imbalance has an important impact on classification performance~\cite{hidaOverviewMachineLearning2026}. In SLR1, of the 126 papers, 61.1\% (77) were excluded and 38.9\% (49) were included in the original study; thus, the accuracy of a model that classifies all papers as excluded would be 0.611. The Llama-3.2-3B-Instruct-Turbo and claude-3-haiku-20240307 models not only reach this baseline but may be close to it, suggesting limited ability to discriminate between classes. For SLR2, of the 392 papers, 67.6\% (265) were excluded and 32.4\% (127) were included; the corresponding accuracy would be 0.676, and in this SLR2 all models from the Llama family presented performance close to the baseline. In contrast, gpt-3.5-turbo achieved an accuracy of 0.37, below the trivial baseline for SLR2. This behavior is consistent with a prediction pattern that is strongly misaligned with the reference labels.

 \begin{figure}[!t]
  \centering
  \subfloat[SLR1]{%
    \includegraphics[width=0.95\columnwidth]{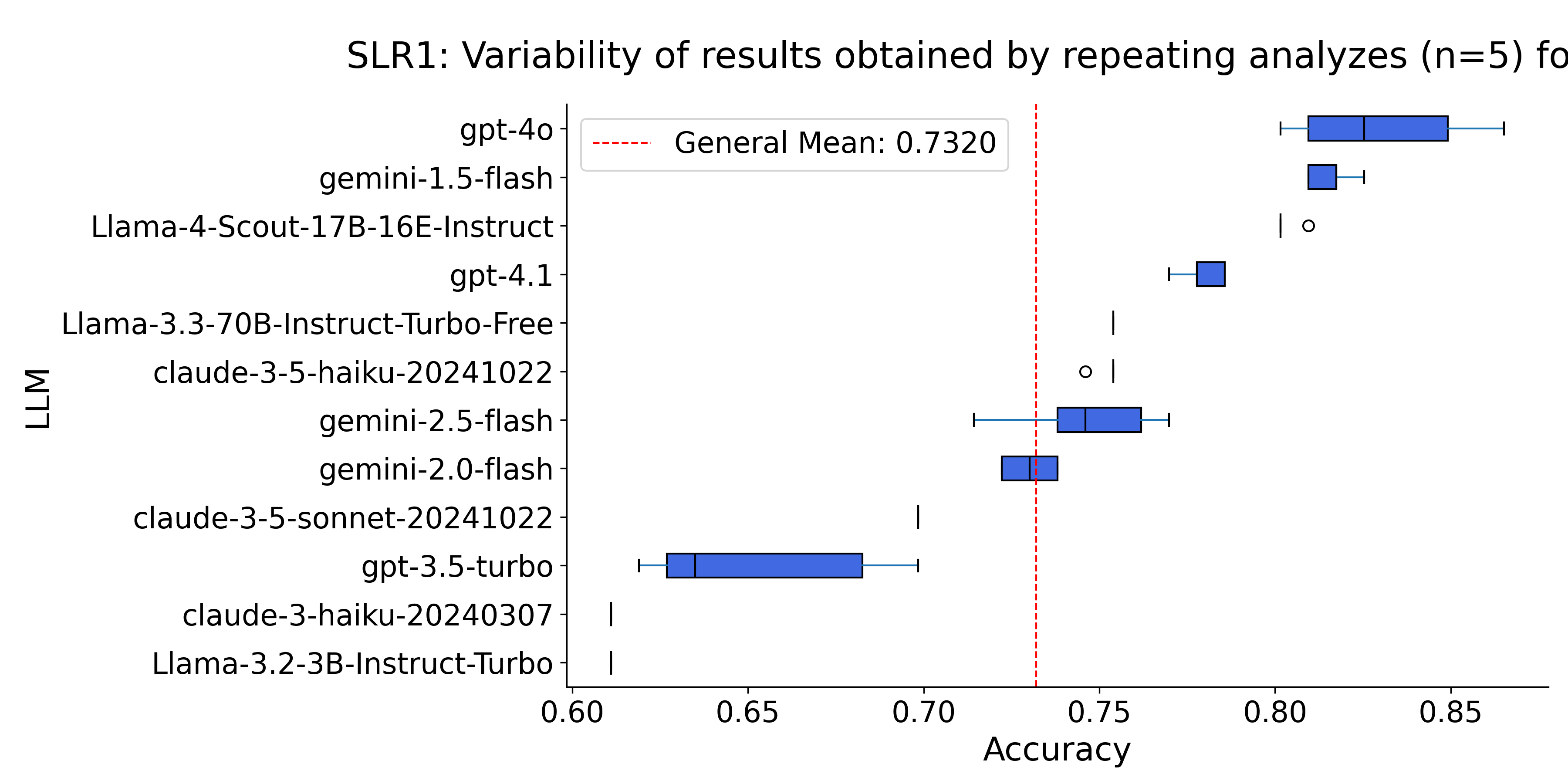}
  }\\[0.3cm]
  \subfloat[SLR2]{%
    \includegraphics[width=0.95\columnwidth]{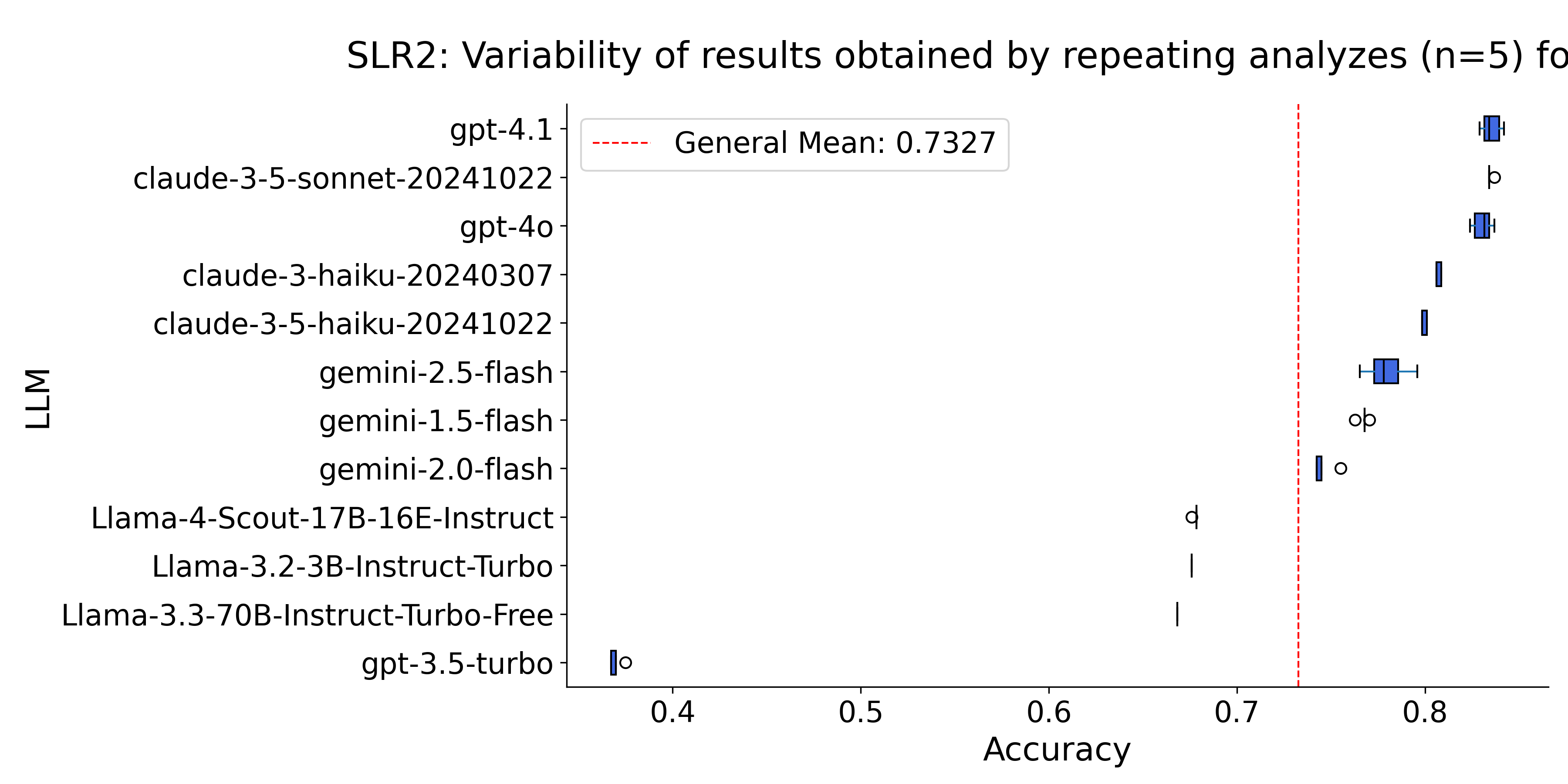}
  }
  \caption{Variability of responses in each iteration (n = 5) per model for both SLR1 and SLR2.}
  \label{fig:slr1_2_variabilidade_modelos}
\end{figure}

\textbf{Main Finding:} Even with temperature set to zero, a substantial portion of the models exhibited response variability across the 5 experimental rounds. Figure~\ref{fig:slr1_2_variabilidade_modelos}  illustrates this behavior. Although the literature acknowledges that zero temperature does not guarantee full determinism in LLMs, this configuration is frequently adopted in empirical screening applications in an attempt to reduce response variation. The empirical results document that, even under this established configuration, multiple iterations reveal inconsistencies in screening decisions.

\begin{figure*}[t]
  \centering

  \subfloat[SLR1-CI1]{%
    \includegraphics[width=0.48\textwidth]{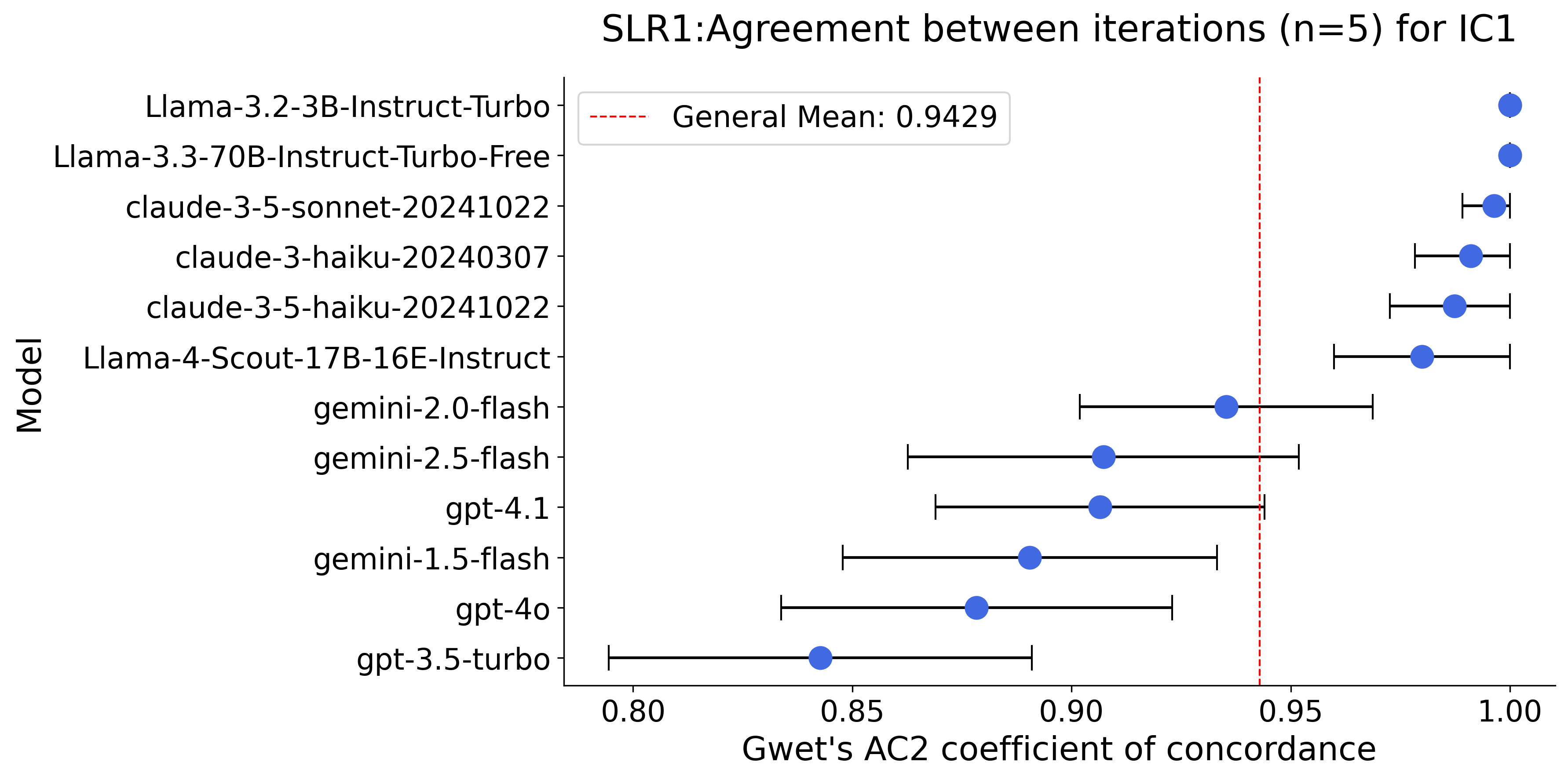}%
    \label{fig:slr1_concordancia_ci1}%
  }\hfill
  \subfloat[SLR1-CI2]{%
    \includegraphics[width=0.48\textwidth]{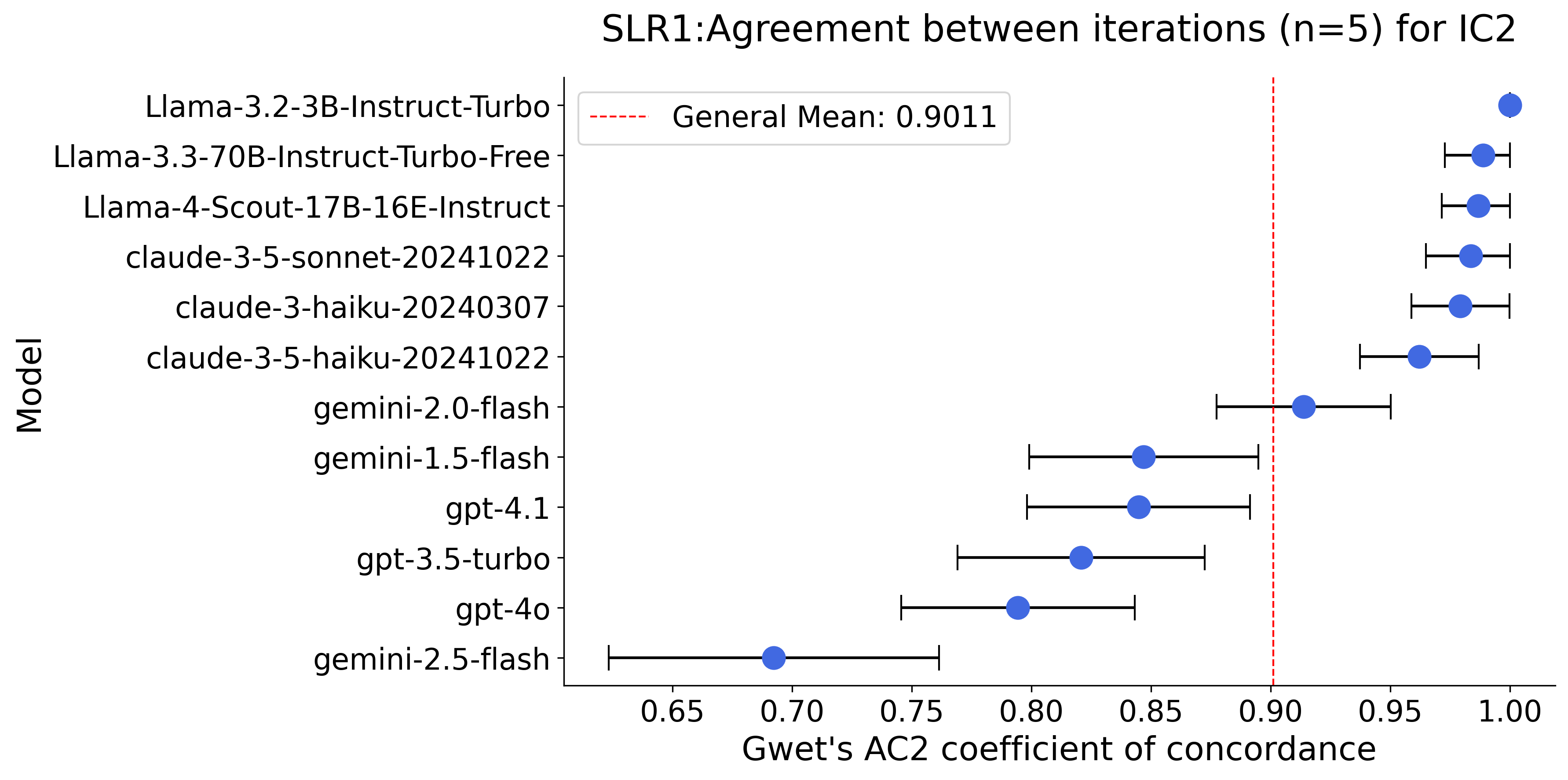}%
    \label{fig:slr1_concordancia_ci2}%
  }\\[\smallskipamount]

  \subfloat[SLR2-CI1]{%
    \includegraphics[width=0.48\textwidth]{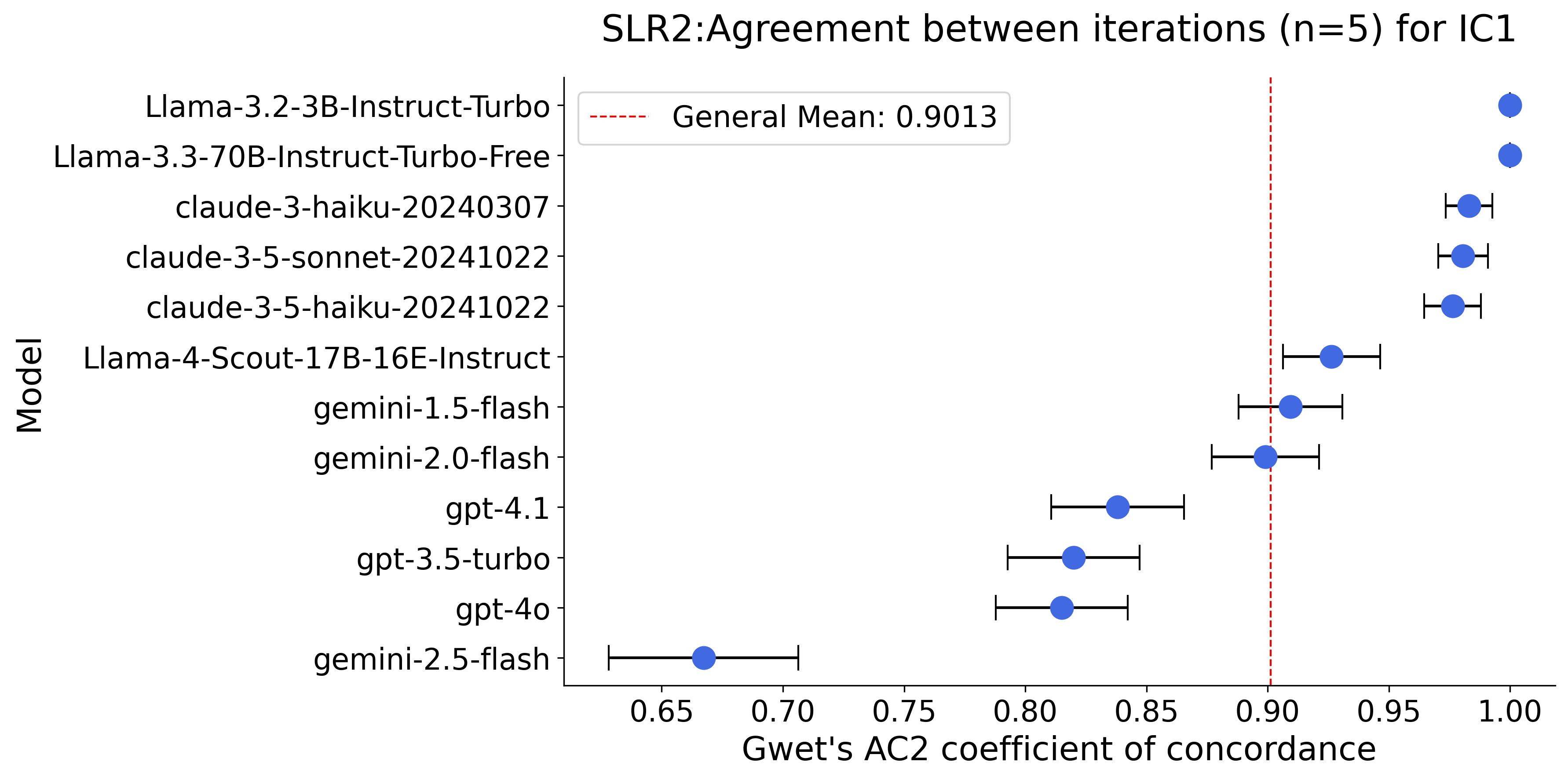}%
  }\hfill
  \subfloat[SLR2-CI2]{%
    \includegraphics[width=0.48\textwidth]{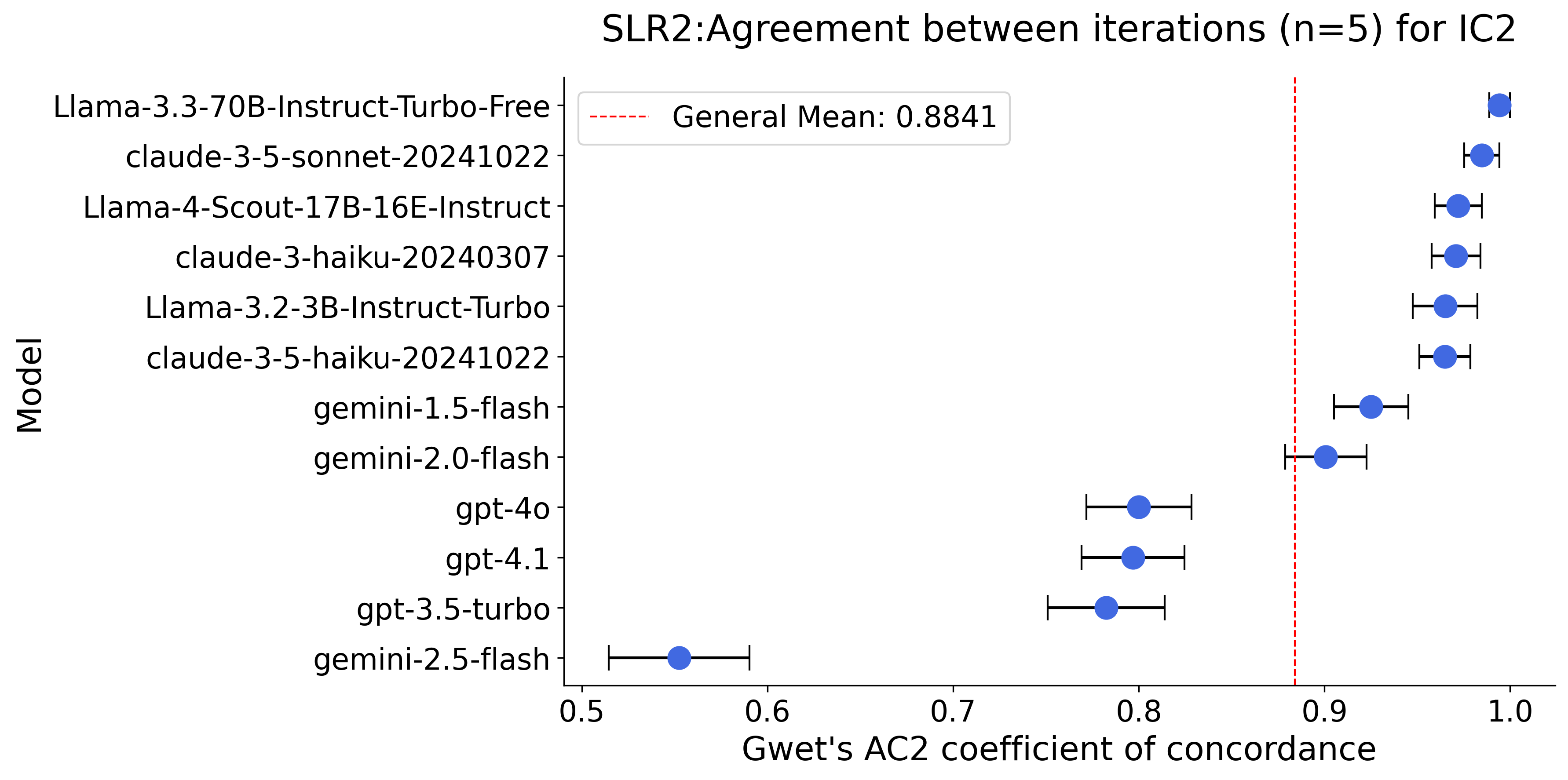}%
    \label{fig:slr2_concordancia_ci2}%
  }

  \caption{Agreement between the responses for each model and SLR across iterations.}
  \label{fig:slr1_2_concordancia}
\end{figure*}

\textbf{Agreement between Iterations}: The analysis using the Gwet AC2 coefficient showed heterogeneity between models (Figure~\ref{fig:slr1_2_concordancia}), with values ranging from 0.55 (Gemini-2.5-Flash) to 1.0 (Llama-3.3-70B-Instruct-Turbo-Free). The agreement varied according to the inclusion criterion and the origin of the studies, suggesting that these characteristics modulate the stability of the response.

\subsection{Phase 2: Impact of Input Features on LLMs}
\label{subsec:resultados_fase2}
We first present results stratified by dataset (SLR1 and SLR2), followed by an aggregated meta-analytic synthesis to assess whether the observed patterns generalize robustly.

\begin{figure}[!t]
  \centering
  \includegraphics[width=0.5\textwidth]{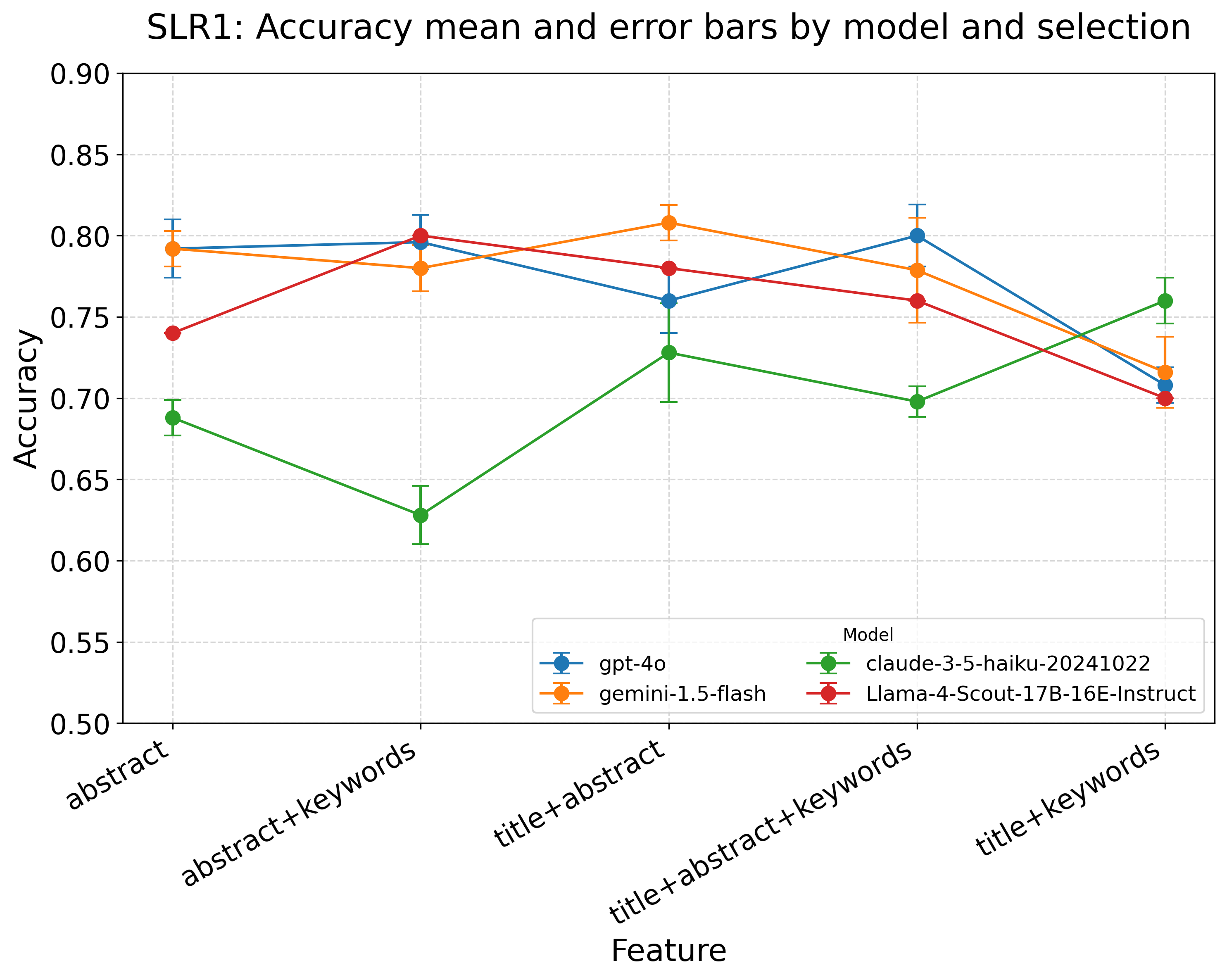} 
  \caption{Phase 2 results for SLR1}
  \label{fig:slr1_resultados_fase2}
\end{figure}

\textbf{SLR1:} We selected 4 models with superior performance in Phase 1: gpt-4o, gemini-1.5-flash, claude-3-5-haiku-20241022 and Llama-4-Scout-17B-16E-Instruct. Figure~\ref{fig:slr1_resultados_fase2} presents the mean accuracy with confidence intervals for each variant of feature composition. Visually, substantial overlap is observed between the 5 variants (except for claude-3-5-haiku-20241022), suggesting equivalence at the level of stratified data.

\begin{figure}[!t]
  \centering
  \includegraphics[width=0.5\textwidth]{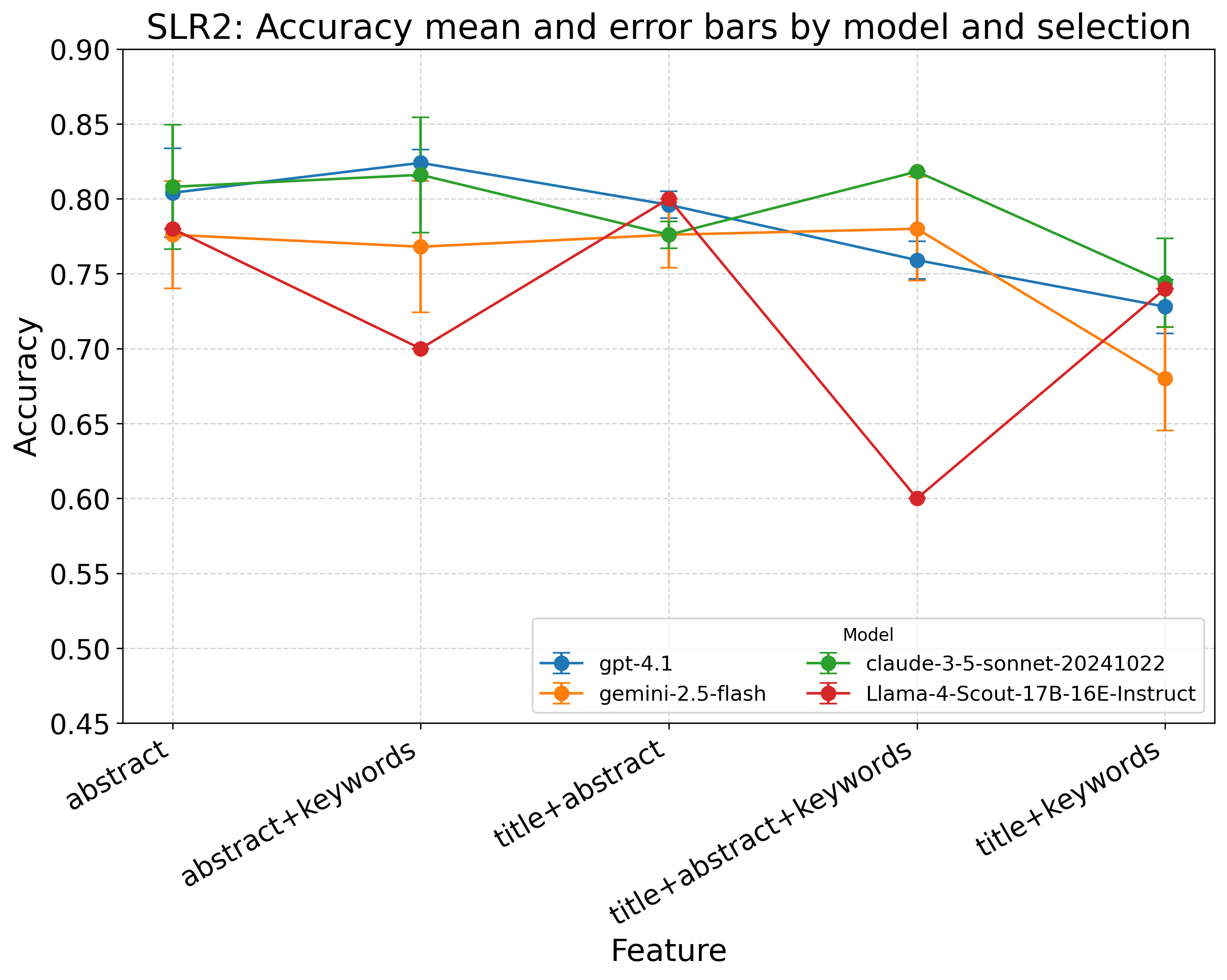} 
  \caption{Phase 2 results for SLR2}
  \label{fig:slr2_resultados_fase2}
\end{figure}

\textbf{SLR2:} The LLMs used were: gpt-4.1, gemini-2.5-flash, claude-3-5-sonnet-20241022 and Llama-4-Scout-17B-16E-Instruct. Figure~\ref{fig:slr2_resultados_fase2} shows a pattern similar to that observed in SLR1, except for Llama-4-Scout-17B-16E-Instruct, which exhibits notably higher variability than the other models.

\begin{figure}[!t]
  \centering
  \includegraphics[width=0.5\textwidth]{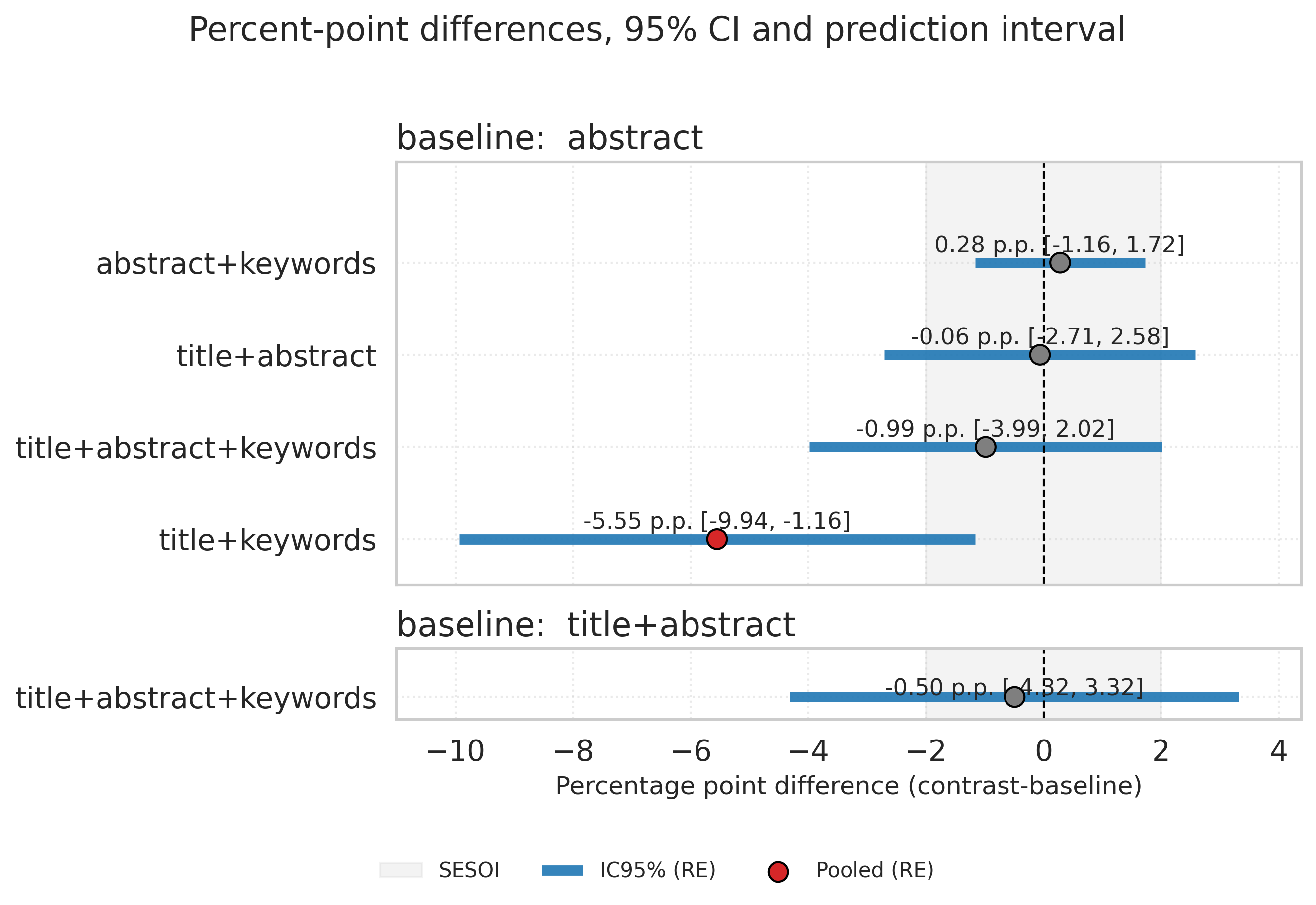} 
  \caption{Meta-analysis of aggregated effect sizes for feature composition variants across both SLRs. Points represent pooled effect estimates; error 
bars show 95\% confidence intervals. Shaded area indicates SESOI (Smallest Effect Size of Interest): ±2.0 p.p.}
  \label{fig:slr1_2_resultados_fase2_meta}
\end{figure}

\textbf{Joint analysis:} To assess whether the patterns observed in each data set were generalized, we conducted a random-effects meta-analysis comparing each feature composition with Variant A (abstract-only metadata). Figure~\ref{fig:slr1_2_resultados_fase2_meta} presents the estimated effects with 95\% confidence intervals.

\begin{itemize}
    \item Variants B, C, and D yielded estimated effects close to zero (abstract+keywords: +0.28 p.p.; title+abstract: -0.06 p.p.; title+abstract+keywords: 
-0.99 p.p.), with 95\% confidence intervals entirely contained within the SESOI of ±2.0 p.p., indicating practical equivalence.
    \item Variant E (title+keywords) produced a significant degradation of -5.55 p.p. (95\% CI: [-9.94, -1.16]), with an interval that does not cross zero and exceeds the SESOI, indicating a meaningful loss.
\end{itemize}

Thus, the presence of an abstract was essential for robust LLM performance. When available, abstracts alone (Variant A) or combined with the title 
and/or keywords (Variants B, C, D) produced equivalent results. Only the composition based exclusively on title and keywords (Variant E) exhibited 
systematically inferior performance, suggesting that LLMs depend on the contextual information provided by the abstract for effective classification.
            
\subsection{Phase 3: Comparison between LLMs and Classical Methods (RQ3)}
\label{subsec:resultados_fase3}
The results of the comparison between LLMs and classical supervised classification methods showed similar performance behaviors between these approaches in the 2 SLRs evaluated.

\begin{figure}[!t]
  \centering
  \includegraphics[width=0.5\textwidth]{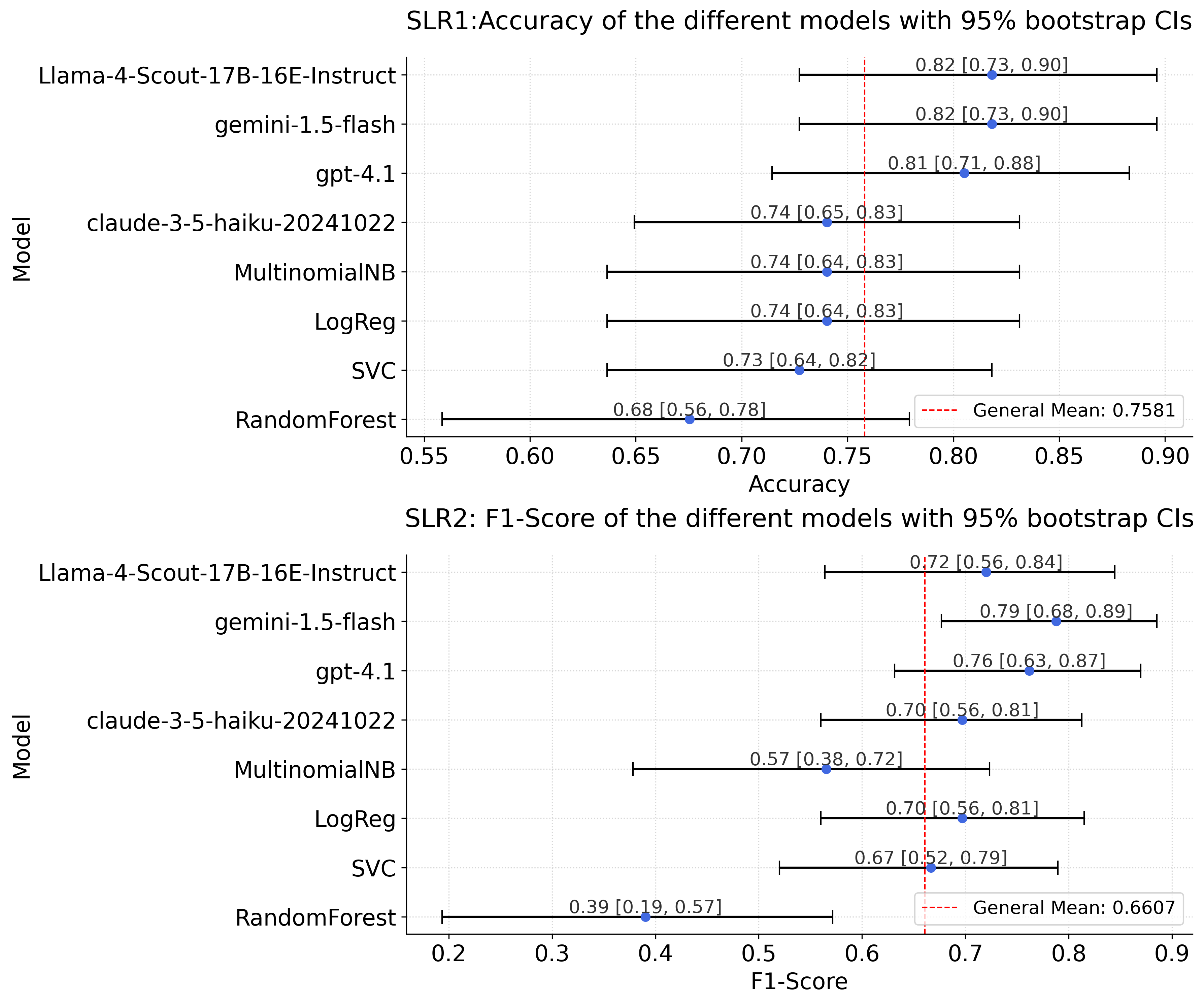} 
  \caption{Comparison of accuracy and F1-score for LLMs and classical machine learning methods on SLR1 with 95\% bootstrap confidence intervals.}
  \label{fig:slr1_fase_3_acc_f1}
\end{figure}

\textbf{Analysis of SLR1 results:} The data presented in Figure~\ref{fig:slr1_fase_3_acc_f1} indicate that LLMs tend to exhibit performance comparable to classical methods. In terms of accuracy, the LLMs Llama-4-Scout-17B-16E-Instruct, gemini-1.5-flash, and gpt-4.1 are concentrated in the range $0.81$-$0.82$, while the classical methods (MultinomialNB, LogReg, SVC) are around $0.74$. Random Forest shows a more pronounced degradation. The estimated confidence intervals exhibit considerable overlap between approaches, which suggests a lack of robust evidence of performance separation under the adopted protocol; therefore, isolated differences should be interpreted with caution.

The F1-score metric reveals more pronounced dynamics. The LLMs (gemini-1.5-flash, gpt-4.1, claude-3-5-haiku) maintained robust performance (F1-scores of $0.70$--$0.79$), whereas Random Forest degraded to $0.39$, suggesting that this algorithm fails to balance precision and recall in this context, which is particularly critical in class-imbalanced settings. The remaining models exhibited an intermediate trend.

\begin{figure}[!t]
  \centering
  \includegraphics[width=0.5\textwidth]{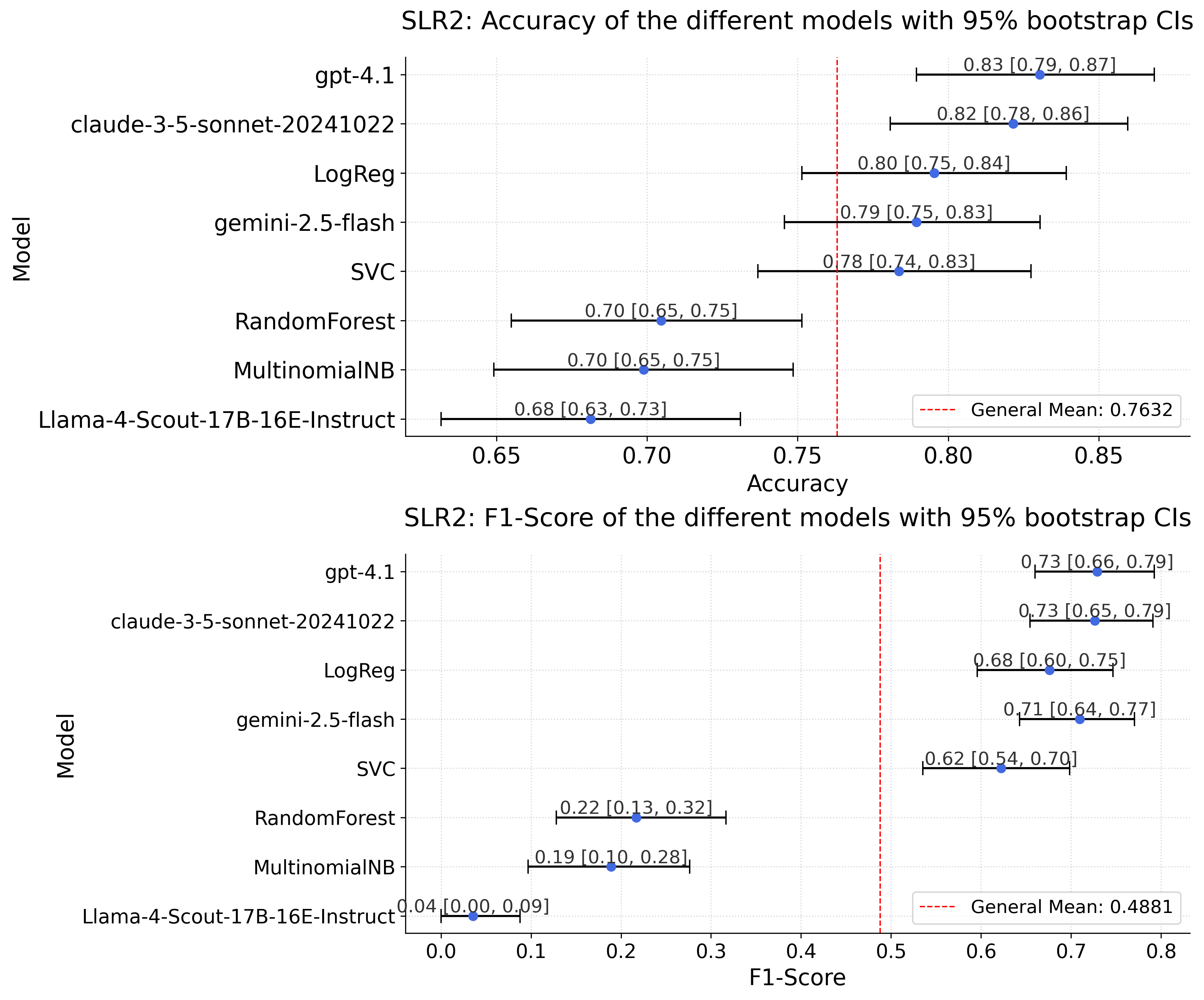} 
  \caption{Comparison of accuracy and F1-score for LLMs and classical machine learning methods on SLR2 with 95\% bootstrap confidence intervals.}
  \label{fig:slr2_fase_3_acc_f1}
\end{figure}

\textbf{Analysis of results for SLR2:} Figure~\ref{fig:slr2_fase_3_acc_f1} presents a pattern similar to SLR1, in which classical methods (particularly Logistic Regression) competed more closely with the LLMs. In terms of accuracy, gpt-4.1 achieved $0.83$, followed by claude-3-5-sonnet-20241022 with $0.82$ and Logistic Regression with $0.80$; thus, Logistic Regression presented a confidence interval that overlaps with those of the LLMs of the top level. An anomalous behavior was detected for Llama-4-Scout-17B-16E-Instruct, which achieved an accuracy of $0.68$ [$0.63$, $0.73$], making it the worst model in SLR2 and constituting a notable reversal compared to its superior performance in SLR1.

For the F1-score, the behaviors are similar. Gpt-4.1 and claude-3-5-sonnet-20241022 achieved F1-scores of $0.73$, while Logistic Regression reached $0.68$, with partially overlapping intervals. This overlap indicates that, statistically, Logistic Regression does not differ significantly from the LLMs on this metric. In contrast, low performance was observed for Random Forest ($0.22$) and Multinomial NB ($0.19$), and particularly for Llama-4-Scout-17B-16E-Instruct ($0.04$). The extremely low F1-score for Llama, approaching zero, confirms a collapse in discrimination: the model assigned virtually all studies to a single class (excluded), resulting in negligible recall. This behavior suggests that, in this specific dataset, the model converged to a degenerate decision strategy, assigning most studies to a single class, which drastically reduces F1-score while preserving part of the accuracy due to class imbalance. The performance reversal of the same model between SLR1 and SLR2 is consistent with the sensitivity to the domain and to the textual distribution of the evaluated set, reinforcing that the results observed in one dataset should not be automatically extrapolated to another.

\textbf{Comparison between SLR1 and SLR2:} Across both SLRs, confidence intervals largely overlapped, supporting broadly comparable performance between LLMs and classical baselines under the adopted protocol. However, domain sensitivity was evident (e.g., Llama-4-Scout reversal from SLR1 to SLR2), and class imbalance can inflate accuracy-based impressions, reinforcing the need to report metrics beyond accuracy when interpreting screening performance.


\section{Operational Guidelines for Adopting LLMs in Study Screening}

This section distills operational guidance from our empirical findings to support a controlled and auditable adoption of LLMs for screening in systematic reviews. The guidance is not normative and should be adapted to each review protocol. The goal is to mitigate risks observed in our experiment, particularly response variability, class imbalance, and metadata constraints, through minimal checks and conservative use.

\subsection{Guidelines Based on the Findings}

A pragmatic starting point is a pilot on a representative sample (e.g., 50--100 studies), comparing at least one LLM with a classical baseline (e.g., Logistic Regression) using multiple metrics selected according to expected class imbalance. Reporting a trivial baseline that excludes all studies helps detect degenerate behaviors and contextualize observed gains.

Metadata availability is a primary constraint: abstracts should be prioritized whenever possible, and LLM-based screening should be avoided when most records lack abstracts. Titles and keywords are complementary signals but should not be treated as substitutes for abstracts.

\begin{figure}[!t]
  \centering
  \includegraphics[width=0.5\textwidth]{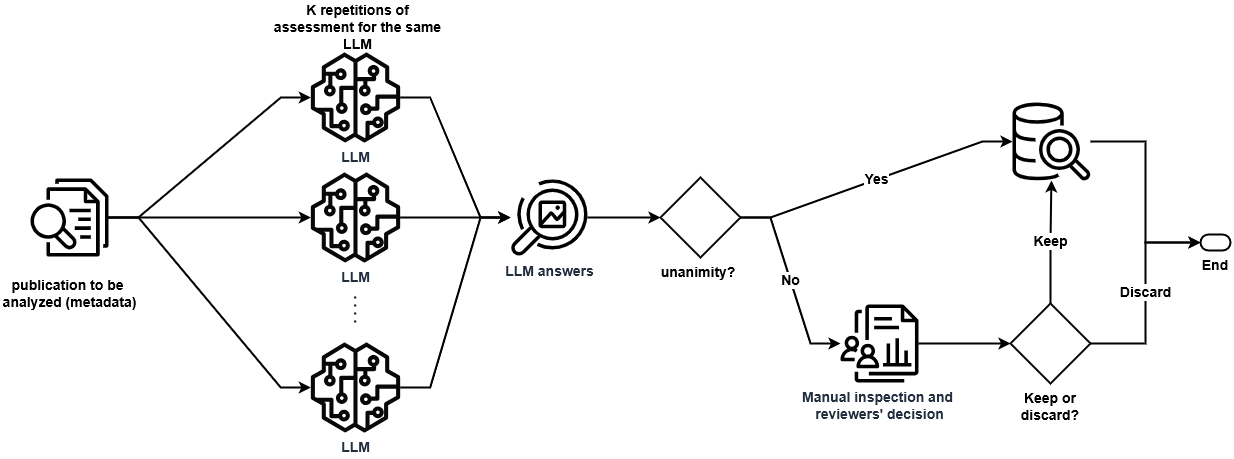} 
  \caption{Flowchart describing our proposed use of LLM in screening; each publication should go through the process described above.}
  \label{fig:diagrama_llm_in_the_loop}
\end{figure}

To account for residual stochasticity, we recommend multi-round inference with unanimity-based automation (Figure~\ref{fig:diagrama_llm_in_the_loop}). Only unanimous decisions are automated; any disagreement flags the study as ``conflicting'' and routes it to human review. In addition, verification sampling on unanimously decided cases helps detect systematic errors due to prompt design or changes in the execution environment.

\subsection{Practical Checklist for Responsible Use}

Before adopting LLMs (or classical models) for SLR screening, we recommend verifying, at a minimum, the following items:

\begin{itemize}
  \item \textbf{Validation:} a pilot study was conducted; at least one LLM and one classical baseline were compared; a trivial exclusion baseline was computed and reported.
  \item \textbf{Metadata:} metadata availability and consistency were verified, especially abstracts; title+keywords-only configurations were avoided when abstracts are not consistently available.
  \item \textbf{Variability:} multiple executions per study were performed; a clear aggregation rule was defined (e.g., unanimity, majority vote, threshold); agreement across runs was quantified and recorded.
  \item \textbf{Evaluation and reproducibility:} performance was reported with multiple metrics (at least precision, recall, and F1); verification sampling was applied to automatically screened cases; software/library versions, environment configuration, inference parameters, and exact model/API identifiers were documented for reproducibility.
\end{itemize}

This checklist is not exhaustive, but it summarizes a minimal set of practices directly supported by our findings, and provides a pragmatic starting point for responsible adoption of LLMs in systematic review screening.

\section{Limitations and Biases}

Although conducted with methodological rigor, this study has scope and design limitations that affect generalizability. This section contextualizes these constraints.

\begin{itemize}
    \item{\textbf{External validity:}} Results are limited to 2 SLRs in educational domains (HCI--AI, Educational Gamification) published in English. Patterns may not generalize to other Software Engineering domains or multilingual corpora.
    
    \item{\textbf{Internal validity:}} Models evaluated via APIs (May--October 2025) introduce temporal dependence; subsequent updates may affect reproducibility.
    
    \item{\textbf{Construct validity:}} Variability estimated from 5 iterations per model suffices to evidence non-determinism but may underestimate variance in unstable settings.
    
    \item{\textbf{Conclusion validity:}} LLM sample excluded fine-tuned or specialized models; no cost--benefit analysis was performed, limiting prescriptive recommendations in resource-constrained contexts. Screening error may partly reflect human ambiguity rather than exclusively model failures.
\end{itemize}

These limitations contextualize the scope without invalidating the findings.


\section{Conclusion}

This study evaluated 12 LLMs from 4 providers and 4 classical machine learning models for screening in 2 real SLRs (n = 518), focusing on cross-model performance heterogeneity, intra-model variability under controlled conditions, and dependence on available metadata.

In Phase 1, we observed substantial heterogeneity across models and sensitivity to the review domain, including rank reversals between SLRs. Even with temperature set to zero, several LLMs produced non-deterministic screening decisions, indicating that stability should be treated as an empirical, context-dependent requirement rather than an operational assumption.

In Phase 2, abstracts emerged as the decisive metadata element for LLM-based screening: removing abstracts consistently degraded performance, whereas adding titles and keywords to abstracts provided no robust practical gains.

In Phase 3, LLMs and classical methods exhibited broadly comparable performance under the same protocol, with overlapping bootstrap confidence intervals. Therefore, method selection should be guided primarily by operational and scientific governance criteria, including cost, transparency, reproducibility, and metadata availability, rather than by aggregated performance metrics alone.

Overall, the study clarifies practical limits and conditions for using LLMs in screening, showing that model scale and version do not guarantee reliability, that residual variability can persist under controlled settings, and that abstract availability is critical. Future work should expand to additional Software Engineering domains and multilingual corpora, strengthen reproducibility analyses in API-based settings by tracking versions and execution windows, increase the number of repeated runs to better characterize variability, and incorporate cost–benefit analyses at scale alongside assessments of reference-label reliability.


\bibliographystyle{unsrtnat}

\bibliography{references}

@misc{costalongaCanMachineLearning2025a,
  title = {Can {{Machine Learning Support}} the {{Selection}} of {{Studies}} for {{Systematic Literature Review Updates}}?},
  author = {Costalonga, Marcelo and Napoleão, Bianca Minetto and Baldassarre, Maria Teresa and Felizardo, Katia Romero and Steinmacher, Igor and Kalinowski, Marcos},
  date = {2025-02-12},
  year  = {2025},
  eprint = {2502.08050},
  eprinttype = {arXiv},
  eprintclass = {cs},
  doi = {10.48550/arXiv.2502.08050},
  url = {http://arxiv.org/abs/2502.08050},
  urldate = {2025-10-16},
  langid = {english},
  pubstate = {prepublished}
}

@misc{espositoGenerativeAIEvidenceBased2024,
  title = {Generative {{AI}} in {{Evidence-Based Software Engineering}}: {{A White Paper}}},
  shorttitle = {Generative {{AI}} in {{Evidence-Based Software Engineering}}},
  author = {Esposito, Matteo and Janes, Andrea and Taibi, Davide and Lenarduzzi, Valentina},
  date = {2024-08-22},
  year  = {2024},
  eprint = {2407.17440},
  eprinttype = {arXiv},
  eprintclass = {cs},
  doi = {10.48550/arXiv.2407.17440},
  url = {http://arxiv.org/abs/2407.17440},
  urldate = {2025-10-16},
  langid = {english},
  pubstate = {prepublished}
}

@inproceedings{felizardoChatGPTApplicationSystematic2024a,
  title = {{{ChatGPT}} Application in {{Systematic Literature Reviews}} in {{Software Engineering}}: An Evaluation of Its Accuracy to Support the Selection Activity},
  shorttitle = {{{ChatGPT}} Application in {{Systematic Literature Reviews}} in {{Software Engineering}}},
  booktitle = {Proceedings of the 18th {{ACM}}/{{IEEE International Symposium}} on {{Empirical Software Engineering}} and {{Measurement}}},
  author = {Felizardo, Katia Romero and Lima, Márcia Sampaio and Deizepe, Anderson and Conte, Tayana Uchôa and Steinmacher, Igor},
  date = {2024-10-24},
  year  = {2024},
  pages = {25--36},
  publisher = {ACM},
  address = {New York, NY, USA},
  location = {Barcelona Spain},
  doi = {10.1145/3674805.3686666},
  url = {https://dl.acm.org/doi/10.1145/3674805.3686666},
  urldate = {2025-10-16},
  eventtitle = {{{ESEM}} '24: {{ACM}} / {{IEEE International Symposium}} on {{Empirical Software Engineering}} and {{Measurement}}},
  isbn = {979-8-4007-1047-6},
  langid = {english}
}

@inproceedings{minettonapoleaoEmergingResultsAutomated2024,
  title = {Emerging {{Results}} on {{Automated Support}} for {{Searching}} and {{Selecting Evidence}} for {{Systematic Literature Review Updates}}},
  booktitle = {Proceedings of the 1st {{IEEE}}/{{ACM International Workshop}} on {{Methodological Issues}} with {{Empirical Studies}} in {{Software Engineering}}},
  author = {Minetto Napoleão, Bianca and Sarkar, Ritika and Hallé, Sylvain and Petrillo, Fabio and Kalinowski, Marcos},
  date = {2024-04-16},
  year  = {2024},
  pages = {34--41},
  publisher = {ACM},
  location = {Lisbon Portugal},
  doi = {10.1145/3643664.3648202},
  url = {https://dl.acm.org/doi/10.1145/3643664.3648202},
  urldate = {2025-10-16},
  eventtitle = {{{WSESE}} '24: 1st {{IEEE}}/{{ACM International Workshop}} on {{Methodological Issues}} with {{Empirical Studies}} in {{Software Engineering}}},
  isbn = {979-8-4007-0567-0},
  langid = {english},
  address   = {New York, NY, USA}
}

@article{syrianiAssessingAbilityChatGPTb,
  title={Screening articles for systematic reviews with ChatGPT},
  author={Syriani, Eugene and David, Istvan and Kumar, Gauransh},
  journal={Journal of Computer Languages},
  volume={80},
  pages={101287},
  year={2024},
  publisher={Elsevier}
}

@article{thodeExploringUseLLMs2025a,
  title = {Exploring the Use of {{LLMs}} for the Selection Phase in Systematic Literature Studies},
  author = {Thode, Lukas and Iftikhar, Umar and Mendez, Daniel},
  date = {2025-08},
  year  = {2025},
  journal = {Information and Software Technology},
  shortjournal = {Information and Software Technology},
  volume = {184},
  pages = {107757},
  issn = {09505849},
  doi = {10.1016/j.infsof.2025.107757},
  url = {https://linkinghub.elsevier.com/retrieve/pii/S0950584925000965},
  urldate = {2025-10-16},
  langid = {english}
}

@article{vandinterAutomationSystematicLiterature2021,
  title = {Automation of Systematic Literature Reviews: {{A}} Systematic Literature Review},
  shorttitle = {Automation of Systematic Literature Reviews},
  author = {Van Dinter, Raymon and Tekinerdogan, Bedir and Catal, Cagatay},
  date = {2021-08},
  year  = {2021},
  journal = {Information and Software Technology},
  shortjournal = {Information and Software Technology},
  volume = {136},
  pages = {106589},
  issn = {09505849},
  doi = {10.1016/j.infsof.2021.106589},
  url = {https://linkinghub.elsevier.com/retrieve/pii/S0950584921000690},
  urldate = {2025-10-16},
  langid = {english}
}

@article{wangErrorRatesHuman2020a,
  title = {Error Rates of Human Reviewers during Abstract Screening in Systematic Reviews},
  author = {Wang, Zhen and Nayfeh, Tarek and Tetzlaff, Jennifer and O’Blenis, Peter and Murad, Mohammad Hassan},
  editor = {Bencharit, Sompop},
  date = {2020-01-14},
  year = {2020},
  journal = {PLOS ONE},
  shortjournal = {PLoS ONE},
  volume = {15},
  number = {1},
  pages = {e0227742},
  issn = {1932-6203},
  doi = {10.1371/journal.pone.0227742},
  url = {https://dx.plos.org/10.1371/journal.pone.0227742},
  urldate = {2025-10-16},
  langid = {english}
}

@misc{brownLanguageModelsAre2020,
  title = {Language {{Models}} Are {{Few-Shot Learners}}},
  author = {Brown, Tom B. and Mann, Benjamin and Ryder, Nick and Subbiah, Melanie and Kaplan, Jared and Dhariwal, Prafulla and Neelakantan, Arvind and Shyam, Pranav and Sastry, Girish and Askell, Amanda and Agarwal, Sandhini and Herbert-Voss, Ariel and Krueger, Gretchen and Henighan, Tom and Child, Rewon and Ramesh, Aditya and Ziegler, Daniel M. and Wu, Jeffrey and Winter, Clemens and Hesse, Christopher and Chen, Mark and Sigler, Eric and Litwin, Mateusz and Gray, Scott and Chess, Benjamin and Clark, Jack and Berner, Christopher and McCandlish, Sam and Radford, Alec and Sutskever, Ilya and Amodei, Dario},
  date = {2020-07-22},
  year  = {2020},
  eprint = {2005.14165},
  eprinttype = {arXiv},
  eprintclass = {cs},
  doi = {10.48550/arXiv.2005.14165},
  url = {http://arxiv.org/abs/2005.14165},
  urldate = {2025-10-17},
  langid = {english},
  pubstate = {prepublished}
}

@misc{huangSurveyHallucinationLarge2025,
  title = {A {{Survey}} on {{Hallucination}} in {{Large Language Models}}: {{Principles}}, {{Taxonomy}}, {{Challenges}}, and {{Open Questions}}},
  shorttitle = {A {{Survey}} on {{Hallucination}} in {{Large Language Models}}},
  author = {Huang, Lei and Yu, Weijiang and Ma, Weitao and Zhong, Weihong and Feng, Zhangyin and Wang, Haotian and Chen, Qianglong and Peng, Weihua and Feng, Xiaocheng and Qin, Bing and Liu, Ting},
  date = {2025-03-31},
  year = {2025},
  journal = {ACM Transactions on Information Systems},
  shortjournal = {ACM Trans. Inf. Syst.},
  volume = {43},
  number = {2},
  eprint = {2311.05232},
  eprinttype = {arXiv},
  eprintclass = {cs},
  numpages = {55},
  issn = {1046-8188, 1558-2868},
  doi = {10.1145/3703155},
  url = {http://arxiv.org/abs/2311.05232},
  urldate = {2025-10-17},
  langid = {english}
}

@misc{xuHallucinationInevitableInnate2025,
  title = {Hallucination Is {{Inevitable}}: {{An Innate Limitation}} of {{Large Language Models}}},
  shorttitle = {Hallucination Is {{Inevitable}}},
  author = {Xu, Ziwei and Jain, Sanjay and Kankanhalli, Mohan},
  date = {2025-02-13},
  year= {2025},
  eprint = {2401.11817},
  eprinttype = {arXiv},
  eprintclass = {cs},
  doi = {10.48550/arXiv.2401.11817},
  url = {http://arxiv.org/abs/2401.11817},
  urldate = {2025-10-17},
  langid = {english},
  pubstate = {prepublished}
}

@misc{rawteSurveyHallucinationLarge2023,
  title = {A {{Survey}} of {{Hallucination}} in {{Large Foundation Models}}},
  author = {Rawte, Vipula and Sheth, Amit and Das, Amitava},
  date = {2023-09-12},
  year = {2023},
  eprint = {2309.05922},
  eprinttype = {arXiv},
  eprintclass = {cs},
  doi = {10.48550/arXiv.2309.05922},
  url = {http://arxiv.org/abs/2309.05922},
  urldate = {2025-10-17},
  langid = {english},
  pubstate = {prepublished}
}

@misc{atilNonDeterminismDeterministicLLM2025,
  title = {Non-{{Determinism}} of "{{Deterministic}}" {{LLM Settings}}},
  author = {Atil, Berk and Aykent, Sarp and Chittams, Alexa and Fu, Lisheng and Passonneau, Rebecca J. and Radcliffe, Evan and Rajagopal, Guru Rajan and Sloan, Adam and Tudrej, Tomasz and Ture, Ferhan and Wu, Zhe and Xu, Lixinyu and Baldwin, Breck},
  date = {2025-04-02},
  year = {2025},
  eprint = {2408.04667},
  eprinttype = {arXiv},
  eprintclass = {cs},
  doi = {10.48550/arXiv.2408.04667},
  url = {http://arxiv.org/abs/2408.04667},
  urldate = {2025-10-27},
  langid = {english},
  pubstate = {prepublished}
}

@inproceedings{renzeEffectSamplingTemperature2024,
  title={The effect of sampling temperature on problem solving in large language models},
  author={Renze, Matthew},
  booktitle={Findings of the association for computational linguistics: EMNLP 2024},
  pages={7346--7356},
  year={2024},
publisher = {Association for Computational Linguistics},
  address   = {Miami, Florida, USA},
}

@book{travassosTertiaryStudyConvergence2017,
  author = {Travassos, Guilherme Horta and Felizardo, Katia Romero and Morandini, Marcelo and Kolski, Christophe},
  title = {A Tertiary Study on the Convergence of Human-Computer Interaction and Artificial Intelligence},
  series = {Learning and Analytics in Intelligent Systems},
  publisher = {Springer International Publishing},
  year = {2017},
  address = {New York, NY, USA},
  edition = {1}
}

@article{pessoaJourneyIdentifyUsers2024,
  title = {A {{Journey}} to {{Identify Users}}' {{Classification Strategies}} to {{Customize Game-Based}} and {{Gamified Learning Environments}}},
  author = {Pessoa, Marcela and Lima, Márcia and Pires, Fernanda and Haydar, Gabriel and Melo, Rafaela and Rodrigues, Luiz and Oliveira, David and Oliveira, Elaine and Galvão, Leandro and Gadelha, Bruno and Isotani, Seiji and Gasparini, Isabela and Conte, Tayana},
  year = {2024},
  journal = {IEEE Transactions on Learning Technologies},
  shortjournal = {IEEE Trans. Learning Technol.},
  volume = {17},
  pages = {527--541},
  issn = {1939-1382, 2372-0050},
  doi = {10.1109/TLT.2023.3317396},
  url = {https://ieeexplore.ieee.org/document/10256042/},
  urldate = {2025-11-03},
  langid = {english}
}

@inproceedings{kitchenham2004evidence,
  title={Evidence-based software engineering},
  author={Kitchenham, Barbara A and Dyba, Tore and Jorgensen, Magne},
  booktitle={Proceedings. 26th International Conference on Software Engineering},
  pages={273--281},
  year={2004},
  organization={IEEE},
  publisher = {IEEE Computer Society},
  address   = {Los Alamitos, CA, USA}
}

@misc{openaiGPTModels2025,
  author = {OpenAI},
  title = {OpenAI API Documentation - GPT Models},
  year = {2025},
  howpublished = {\url{https://platform.openai.com/docs/models}},
  note = {Accessed on: 2025-11-03},
  organization = {OpenAI}
}

@misc{googleGeminiModels2025,
  author = {Google DeepMind},
  title = {Gemini API Documentation},
  year = {2025},
  howpublished = {\url{https://ai.google.dev/docs/models}},
  note = {Accessed on: 2025-11-03},
  organization = {Google DeepMind}
}

@misc{anthropicClaudeModels2025,
  author = {Anthropic},
  title = {Claude API Documentation},
  year = {2025},
  howpublished = {\url{https://docs.anthropic.com/en/resources/overview}},
  note = {Accessed on: 2025-11-03},
  organization = {Anthropic}
}

@misc{togetherai2025,
  author = {Together AI},
  title = {Together AI Platform - Llama Models},
  year = {2025},
  howpublished = {\url{https://www.together.ai/}},
  note = {Accessed on: 2025-11-03},
  organization = {Together AI}
}

@article{hidaOverviewMachineLearning2026,
  title = {Overview of Machine Learning in Class Imbalance Scenarios: {{Trends}}, Challenges, and Approaches},
  shorttitle = {Overview of Machine Learning in Class Imbalance Scenarios},
  author = {Hida, Gilberto Sussumu and Alves Do Nascimento, André Câmara},
  date = {2026-03},
  year={2026},
  journal = {Expert Systems with Applications},
  shortjournal = {Expert Systems with Applications},
  volume = {298},
  pages = {129592},
  issn = {09574174},
  doi = {10.1016/j.eswa.2025.129592},
  url = {https://linkinghub.elsevier.com/retrieve/pii/S0957417425032075},
  urldate = {2025-11-06},
  langid = {english}
}

@article{dersimonianMetaAnalysisClinicalTrials2015,
  author = {DerSimonian, Rebecca and Laird, Nan M.},
  title = {Meta-Analysis in Clinical Trials Revisited},
  journal = {Contemporary Clinical Trials},
  volume = {45},
  pages = {139--145},
  year = {2015},
  doi = {10.1016/j.cct.2015.09.002},
  pmid = {26343745}
}

@article{tibshirani1993introduction,
  title={An introduction to the bootstrap},
  author={Tibshirani, Robert J and Efron, Bradley},
  journal={Monographs on statistics and applied probability},
  volume={57},
  number={1},
  pages={1--436},
  year={1993}
}

@article{anvari2021using,
  title={Using anchor-based methods to determine the smallest effect size of interest},
  author={Anvari, Farid and Lakens, Dani{\"e}l},
  journal={Journal of Experimental Social Psychology},
  volume={96},
  pages={104159},
  year={2021},
  publisher={Elsevier}
}

@book{bishop2006pattern,
  author = {Bishop, Christopher M.},
  title = {Pattern Recognition and Machine Learning},
  year = {2006},
  publisher = {Springer},
  address = {New York, NY},
  isbn = {978-0-387-31073-2}
}

@book{hastie2009elements,
  author = {Hastie, Trevor and Tibshirani, Robert and Friedman, Jerome},
  title = {The Elements of Statistical Learning: Data Mining, Inference, and Prediction},
  edition = {2nd},
  year = {2009},
  publisher = {Springer},
  address = {New York, NY},
  isbn = {978-0-387-84857-0}
}

@article{saltonTermWeightingAutomatic1988,
  author = {Salton, Gerard and Buckley, Christopher},
  title = {Term-Weighting Approaches in Automatic Text Retrieval},
  journal = {Information Processing \& Management},
  volume = {24},
  number = {5},
  pages = {513--523},
  year = {1988},
  doi = {10.1016/0306-4573(88)90021-0}
}

@book{bird2009natural,
  author = {Bird, Steven and Klein, Ewan and Loper, Edward},
  title = {Natural Language Processing with Python},
  year = {2009},
  publisher = {O'Reilly Media},
  address = {Sebastopol, CA},
  isbn = {978-0-596-51649-9}
}

@article{pedregosa2011scikit,
  author = {Pedregosa, Fabian and others},
  title = {Scikit-learn: Machine Learning in Python},
  journal = {Journal of Machine Learning Research},
  volume = {12},
  pages = {2825--2830},
  year = {2011}
}

@book{gwet2014handbook,
  title={Handbook of inter-rater reliability: The definitive guide to measuring the extent of agreement among raters},
  author={Gwet, Kilem L},
  year={2014},
  address   = {Gaithersburg, MD, USA},
  publisher={Advanced Analytics, LLC}
}

@article{luo2024potential,
  title={Potential roles of large language models in the production of systematic reviews and meta-analyses},
  author={Luo, Xufei and Chen, Fengxian and Zhu, Di and Wang, Ling and Wang, Zijun and Liu, Hui and Lyu, Meng and Wang, Ye and Wang, Qi and Chen, Yaolong},
  journal={Journal of Medical Internet Research},
  volume={26},
  pages={e56780},
  year={2024},
  publisher={JMIR Publications Toronto, Canada}
}

@article{chen2024chatgpt,
  title={How is ChatGPT’s behavior changing over time?},
  author={Chen, Lingjiao and Zaharia, Matei and Zou, James},
  journal={Harvard Data Science Review},
  volume={6},
  number={2},
  pages   = {1--47},
  year={2024},
  publisher={The MIT Press}
}

@misc{angermeir2025reflections,
  title={Reflections on the Reproducibility of Commercial LLM Performance in Empirical Software Engineering Studies},
  author={Angermeir, Florian and Amougou, Maximilian and Kreitz, Mark and Bauer, Andreas and Linhuber, Matthias and Fucci, Davide and Mendez, Daniel and Gorschek, Tony and others},
  journal={arXiv preprint arXiv:2510.25506},
  year={2025}
}

@inproceedings{bender2021dangers,
  title     = {On the Dangers of Stochastic Parrots: Can Language Models Be Too Big?},
  author    = {Bender, Emily M. and Gebru, Timnit and McMillan-Major, Angelina and Shmitchell, Shmargaret},
  booktitle = {Proceedings of the 2021 ACM Conference on Fairness, Accountability, and Transparency},
  pages     = {610--623},
  year      = {2021},
  publisher = {Association for Computing Machinery},
  address   = {New York, NY, USA},
  doi       = {10.1145/3442188.3445922}
}

@article{cohen2006reducing,
  title={Reducing workload in systematic review preparation using automated citation classification},
  author={Cohen, Aaron M and Hersh, William R and Peterson, Kim and Yen, Po-Yin},
  journal={Journal of the American Medical Informatics Association},
  volume={13},
  number={2},
  pages={206--219},
  year={2006},
  publisher={BMJ Group BMA House, Tavistock Square, London, WC1H 9JR}
}

\end{document}